\documentclass[sigconf, nonacm]{acmart}

\usepackage{pvldb}% VLDB/PVLDB formatting overrides, see pvldb.sty for details.
\usepackage{tikz}
\usepackage{xurl}

\usepackage[table]{xcolor}
\usepackage{booktabs}
\usepackage{tabularx}
\usepackage{array}
\usepackage{xurl}
\usepackage{color}
\usepackage{url}
\usepackage{algorithm}
\usepackage{hyperref}
\usepackage{algpseudocode}
\definecolor{ForestGreen}{rgb}{0.13,0.55,0.13}

\renewcommand\footnotetextcopyrightpermission[1]{}

\usepackage{seqsplit}
\usepackage{caption}

\usepackage{float}
\usepackage{xcolor}
\definecolor{AGreen}{RGB}{35,130,70}
\definecolor{ARed}{RGB}{190,55,45}
\definecolor{ABlue}{RGB}{45,95,180}
\definecolor{AGray}{RGB}{80,80,80}

\usepackage[breakable]{tcolorbox}

\usepackage{enumitem}

\newtcolorbox{answerbox}{
  breakable,
  colback=black!5,
  colframe=black!12,
  boxrule=0.3pt,
  arc=2pt,
  left=3pt,
  right=3pt,
  top=3pt,
  bottom=3pt,
  before skip=3pt,
  after skip=3pt
}

\usepackage{xurl}
\renewcommand\vldbdoi{XX.XX/XXX.XX}
\renewcommand\vldbpages{XXX-XXX}
\renewcommand\vldbavailabilityurl{https://github.com/xinyueyangiscas/AgenticDB}

\begin{document}
\title{AgenticDB: Self-Evolving Reconfiguration Framework for Database Workloads}
%%
%% The "author" command and its associated commands are used to define the authors and their affiliations.
\author{
Xinyue Yang\textsuperscript{1,2},
Chaozheng Wang\textsuperscript{3},
Chen Zheng\textsuperscript{1,2,*},
Heng Zhang\textsuperscript{2},
Yanjun Wu\textsuperscript{2}
}

\affiliation{
\institution{
\textsuperscript{1}University of Chinese Academy of Sciences, Nanjing, China\\
\textsuperscript{2}Institute of Software, Chinese Academy of Sciences, Beijing, China\\
\textsuperscript{3}The Chinese University of Hong Kong, Hong Kong, China
}
\country{}
}

%%
%% The abstract is a short summary of the work to be presented in the
%% article.
\begin{abstract}

Configuration tuning is critical to database performance but remains
difficult in real deployments. Despite notable advances, prior methods
still leave substantial performance potential unexplored, suffer from
low tuning efficiency, and provide limited support for configuration
validation and failure recovery.

To address these limitations, we propose \textsc{AgenticDB}, a
self-evolving agentic framework for database workload reconfiguration.
\textsc{AgenticDB} uses a large language model (LLM)-based DBA Planner to jointly reconfigure database knobs and operating system (OS) parameters through two
key mechanisms. First,
\emph{context-grounded bottleneck diagnosis} uses workload
characteristics, configuration state, and observed runtime behavior to
identify the current performance bottleneck and recommend targeted database management system (DBMS)/OS
reconfiguration actions. Second, \emph{closed-loop context evolution} uses observed performance
and runtime-state changes as feedback to update the bottleneck diagnosis,
guide subsequent decisions, and terminate the reconfiguration loop when
performance plateaus. It also consolidates accumulated reconfiguration experience for reuse on workloads with similar characteristics. Beyond
these two mechanisms, \textsc{AgenticDB} improves reliability by
validating each proposed configuration before applying it and
automatically recovering from failures.

We evaluate \textsc{AgenticDB} on MySQL and PostgreSQL using YCSB,
Sysbench, and TPC-H. Compared with SOTA methods, \textsc{AgenticDB}
outperforms the best-performing baseline by 118.1\% on average and
reduces the total time-to-best across workloads by 22.6\%. Further
analyses show that validation and recovery improve reconfiguration
reliability. The consolidated experience also helps \textsc{AgenticDB} reach high-performing configurations earlier on workloads with similar
characteristics.

\end{abstract}

\maketitle

%%% do not modify the following VLDB block %%
%%% VLDB block start %%%
\vldbtopmatter
%%% VLDB block end %%%

\section{Introduction}

Database configuration tuning is critical for workload performance~\cite{zhang2022towards,zhang2023unified,zhan2024knobtune}. Modern DBMSs expose hundreds of knobs controlling memory management, concurrency, logging, caching, query execution, and I/O behavior~\cite{hpo_db_tuning,too_many_knobs}. Their effects vary substantially across workloads, hardware environments, and interactions with other knobs, resulting in a large and heterogeneous configuration space~\cite{too_many_knobs,sard}. Consequently, manual tuning is difficult to scale and often fails to identify high-performance configurations for a target deployment~\cite{sard,too_many_knobs}.

Existing automatic tuning systems span several representative lines of work. Search- and learning-based methods use Bayesian optimization or reinforcement learning to iteratively explore predefined configuration spaces. To shorten tuning time, they often narrow these spaces through knob selection, dimensionality reduction, value bucketization, or range pruning
~\cite{ituned,cdbtune,qtune,hunter,llamatune,yang2026plrtune}. Historical-data-driven methods usually require substantial time to
collect historical tuning data and train models, and then transfer the learned experience to new workloads~\cite{ottertune,restune,e2etune}. Documentation-guided and LLM-based methods exploit manuals, community knowledge, and language-model reasoning to guide knob-space construction and configuration recommendation~\cite{dbbert,gptuner,lambdatune,agenttune}. Together, these methods have substantially improved the efficiency and effectiveness of database configuration tuning.

Despite these advances, prior tuning methods still face three key
challenges in real deployments.

\textbf{C1. Prior methods leave substantial performance potential unexplored.}
There are three main reasons for this limitation. First, prior methods
face a cost--coverage tradeoff: within a limited tuning budget, it is
difficult to sufficiently explore a broad DBMS knob space and identify
a high-performing configuration. To control tuning cost, they focus
exploration on knobs and values that are expected to have stronger
performance impact. Specifically, they may preselect DBMS
knobs~\cite{hunter,gptuner,agenttune,yang2026plrtune}, bucketize
numerical values~\cite{llamatune}, or prune candidate
ranges~\cite{gptuner,agenttune}. Although these choices reduce tuning
cost, they also reduce configuration coverage: influential parameters
may be excluded, promising values may be skipped, and effective
regions may be narrowed or discarded~\cite{ottertune,cdbtune}.
Consequently, the reduced space may exclude configurations capable of
achieving higher performance, leaving potential gains unrealized for
the target deployment.

Second, prior methods do not explicitly interpret runtime-state changes
to identify precise directions for subsequent exploration
~\cite{cdbtune,agenttune}. MySQL runtime-state sources such as
\texttt{SHOW GLOBAL STATUS} and
\texttt{INFORMATION\_SCHEMA.INNODB\_METRICS}, together with PostgreSQL
statistics views such as \texttt{pg\_stat\_*}, expose buffer behavior,
logging and I/O activity, lock waits, and other execution
behavior~\cite{mysql_status,mysql_innodb_metrics,postgres_monitoring}.
Changes in these states can indicate whether performance is still
limited by buffer behavior, logging overhead, I/O pressure, lock waits,
or other runtime factors, and can guide exploration toward the
corresponding parameters. Although CDBTune~\cite{cdbtune} encodes runtime measurements as database states
for policy learning and AgentTune~\cite{agenttune} provides them as execution features
to its configuration recommender, neither explicitly relates runtime-state changes to bottleneck conditions to guide subsequent exploration. As a result, under a limited tuning budget, prior methods may explore
parameters that are weakly related to the current bottleneck, making it
difficult to reach higher-performing configurations quickly.

Third, prior methods generally operate only over DBMS knobs, without
considering OS-level parameters~\cite{cdbtune,llamatune}. For example,
residual I/O pressure may depend on OS-level storage scheduling even
after relevant DBMS knobs have been tuned, leaving additional
performance gains outside the DBMS-only tuning space
~\cite{redhat_oracle_tuning,io_scheduler_pgbench}.

\textbf{C2. Prior methods suffer from low tuning efficiency.}
Prior methods commonly tune databases through an iterative process
guided by objective measurements. In each round, they apply a candidate
configuration, measure the target objective, such as throughput,
latency, or execution time, and use the result as feedback to generate
the next configuration
~\cite{ituned,ottertune,cdbtune,qtune,hunter,dbbert,gptuner,agenttune}.
Although these methods can refine configurations according to measured
performance, they do not directly explain why performance changes or
whether the current tuning direction can still bring further
improvement. Consequently, they may need many rounds to
find effective configurations and may keep testing low-value candidates
after performance gains have largely saturated. An efficient loop
should therefore analyze performance and runtime-state changes together
to identify the remaining bottleneck and decide whether to continue the
current tuning direction, change to another direction, or terminate the
process, thereby avoiding low-value trials and reducing unnecessary
tuning rounds.

\textbf{C3. Prior methods provide limited configuration validation
and failure recovery.}
Improper values or incompatible parameter combinations may destabilize
the database service or cause configuration failures. These risks are
particularly serious for DBMS knobs that take effect only after a
reload or restart. Invalid settings may be rejected or fail to take
effect when the configuration is reloaded, while incompatible
restart-time settings may prevent the DBMS from restarting
successfully~\cite{postgres_config_setting,postgres_shared_buffers,
agenttune,postgres_wal_config}. For example, overly large buffer-pool
settings may exhaust memory or destabilize the system~\cite{agenttune},
and PostgreSQL cannot start when \texttt{wal\_level=minimal} is combined
with a nonzero \texttt{max\_wal\_senders}
~\cite{postgres_wal_config}.

Prior methods may reduce these risks by tuning only dynamically
changeable knobs, excluding restart-sensitive or potentially harmful
parameters, or restricting candidate value ranges
~\cite{qtune,ottertune}. Such conservative strategies may omit
influential knobs that require a reload or restart to take effect or
exclude promising values for retained knobs, thereby limiting
achievable performance improvements
~\cite{postgres_shared_buffers,mysql_buffer_pool_restart}. Moreover,
these coarse-grained restrictions do not validate each generated
configuration against hardware constraints and cross-parameter
dependencies. Invalid or incompatible configurations may therefore
still be produced, causing changes to be rejected, fail to take effect,
or prevent the DBMS from restarting successfully. An effective tuning
system should strive to generate configurations that satisfy validity
and compatibility constraints, validate each generated configuration
before applying it, and automatically recover if a failure still occurs.

To address these challenges, we propose \textsc{AgenticDB}, a
self-evolving agentic framework for database workload reconfiguration.
At its core, \textsc{AgenticDB} implements a context-grounded harness
that connects an LLM DBA Planner with the target DBMS/OS environment
and organizes reconfiguration as an iterative diagnose--act--observe
process. The harness coordinates DBMS-level and OS-level
reconfiguration, starting with DBMS knobs and switching to OS
parameters when necessary. It is built around two key mechanisms:
\emph{context-grounded bottleneck diagnosis} and
\emph{closed-loop context evolution}.

Through the first mechanism, \emph{context-grounded bottleneck
diagnosis}, the LLM DBA Planner jointly analyzes workload
characteristics, configuration state, and observed runtime behavior to
diagnose the current performance bottleneck and recommend targeted
DBMS/OS reconfiguration actions. Instead of preselecting a small subset
of knobs or pruning value ranges before tuning, \textsc{AgenticDB}
keeps a broad DBMS/OS reconfiguration scope available and uses the
diagnosed bottleneck to recommend the parameter--value changes that are
most relevant in each round. This allows \textsc{AgenticDB} to consider
configurations that prior methods may exclude, without exhaustively
enumerating the entire parameter space. By preserving a broad DBMS/OS scope and recommending parameter--value
changes according to the diagnosed bottleneck, \textsc{AgenticDB} can
uncover performance gains missed by methods that restrict tuning to a
reduced DBMS knob space (addressing C1).

Through the second mechanism, \emph{closed-loop context evolution}, the
LLM DBA Planner uses the observed performance and runtime-state changes
from each round to update the bottleneck diagnosis and guide subsequent
reconfiguration decisions. The outcomes of previous rounds are carried
into the next planning context, allowing subsequent diagnosis and
reconfiguration to focus on unresolved bottlenecks rather than treating
each configuration trial independently. As the loop progresses,
\textsc{AgenticDB} refines the configuration for the current bottleneck
when gains remain, redirects the tuning process toward another
bottleneck when gains from the current one diminish, switches from DBMS
knobs to OS parameters once DBMS-level performance plateaus, and
terminates the loop when no substantial improvement remains.
\textsc{AgenticDB} also consolidates accumulated reconfiguration
experience for reuse on workloads with similar characteristics. By
updating the bottleneck diagnosis from prior outcomes and using it to
guide loop control, this mechanism avoids low-value trials
and reduces unnecessary tuning rounds, thereby improving tuning
efficiency (addressing C2).

\begin{table}[t]
\centering
\captionsetup{skip=2pt}
\caption{Representative Tuning Actions}
\label{tab:motivation-knobs}

\scriptsize
\setlength{\tabcolsep}{3pt}
\renewcommand{\arraystretch}{1.06}

\begin{tabularx}{\columnwidth}{
  @{}
  >{\raggedright\arraybackslash}p{0.12\columnwidth}
  >{\raggedright\arraybackslash}p{0.33\columnwidth}
  @{\hspace{9pt}}
  >{\raggedright\arraybackslash}X
  @{}
}
\toprule
{\footnotesize\textbf{Workload}}
& {\footnotesize\textbf{AgentTune}}
& {\footnotesize\textbf{AgenticDB}} \\
\midrule

PostgreSQL Read
&
\texttt{shared\_buffers}: $[0.125,32]$\,GB $\rightarrow 8$\,GB;
\texttt{random\_page\_cost}: $[1,8]\rightarrow1.1$
&
\texttt{effective\_cache\_size}: $[0.5,51.2]$\,GB $\rightarrow48$\,GB;
\newline
\texttt{plan\_cache\_mode} $\rightarrow\texttt{force\_generic\_plan}$;
\texttt{jit} $\rightarrow\texttt{off}$;
\texttt{max\_parallel\_workers\_per\_gather} $\rightarrow$ tuned
\\
\midrule

MySQL TPC-H
&
\texttt{innodb\_buffer\_pool\_size}: $[0.125,48]$\,GB
$\rightarrow0.125$\,GB;
\texttt{join/sort\_buffer\_size}: $[0.25,64]$\,MB
$\rightarrow0.25$\,MB
&
DBMS: \texttt{innodb\_buffer\_pool\_size} $\rightarrow40$\,GB;
\texttt{innodb\_parallel\_read\_threads} $\rightarrow12$;
\newline
OS: \texttt{block.read\_ahead\_kb} $\rightarrow2048$;
\texttt{block.nr\_requests} $\rightarrow1024$
\\
\bottomrule
\end{tabularx}

% \vspace{-2mm}
\end{table}

\begin{figure}[t]
\centering

\includegraphics[
    width=0.85\columnwidth,
    trim=0mm 2mm 0mm 2mm,
    clip
]{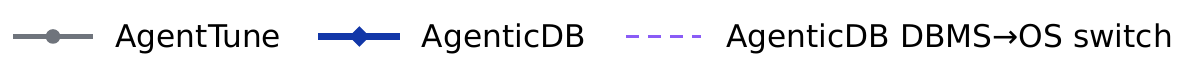}

\includegraphics[
    width=0.485\columnwidth,
    trim=1mm 3mm 1mm 3mm,
    clip
]{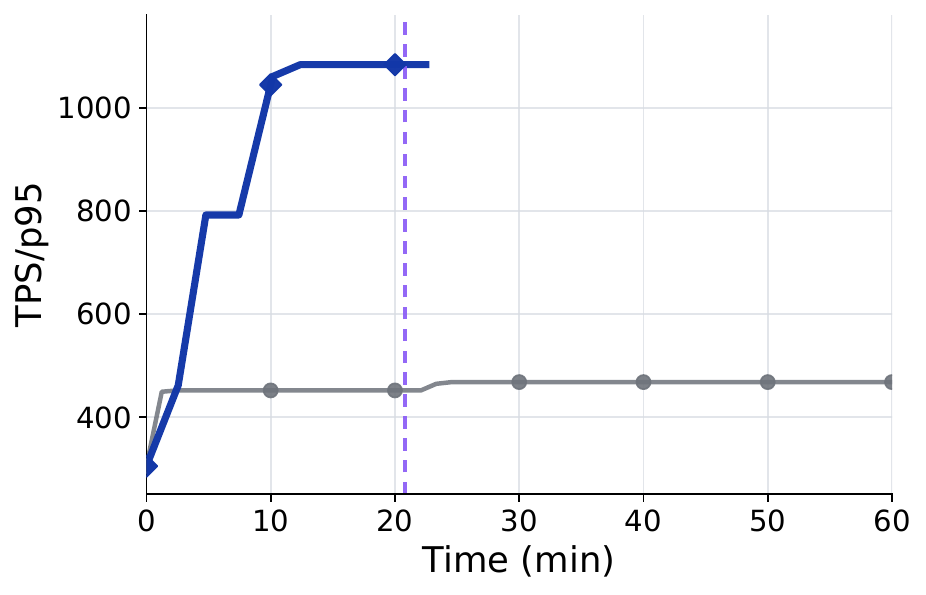}
\hfill
\includegraphics[
    width=0.485\columnwidth,
    trim=1mm 3mm 1mm 3mm,
    clip
]{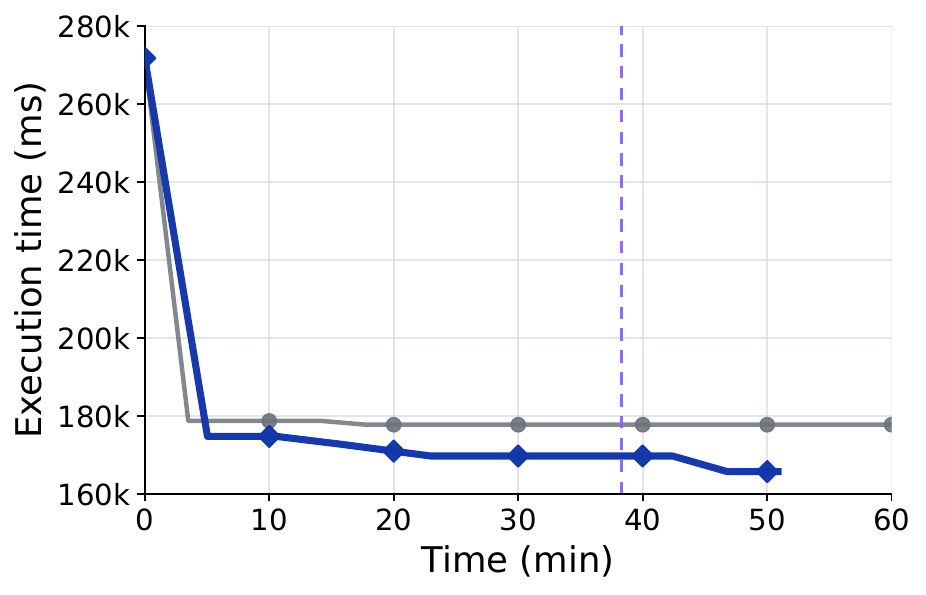}

\makebox[0.485\columnwidth][c]
{\footnotesize\textbf{(a) PostgreSQL Read}}
\hfill
\makebox[0.485\columnwidth][c]
{\footnotesize\textbf{(b) MySQL TPC-H}}
% \captionsetup{skip=2pt}
\caption{Performance comparison between AgentTune and AgenticDB.}
\label{fig:motivation-performance}
\vspace{-3mm}
\end{figure}

Table~\ref{tab:motivation-knobs} and
Figure~\ref{fig:motivation-performance} illustrate how
\textsc{AgenticDB} diagnoses bottlenecks and evolves its
reconfiguration decisions. On PostgreSQL Sysbench Read, runtime-state
signals such as near-zero WAL writes and I/O fsyncs rule out
write-path pressure, while high block reads, evictions, and CPU load
indicate cache churn and query-execution overhead. This diagnosis
guides the LLM DBA Planner toward relevant DBMS knobs beyond
AgentTune's selected subset. Within the displayed 60-minute window,
\textsc{AgenticDB} reaches 1,083.60 TPS/p95 in 12.38 minutes, whereas
AgentTune reaches 467.62 TPS/p95 in 24.57 minutes.

On MySQL TPC-H, high buffer-pool reads and physical data reads indicate
buffer and physical-read pressure, guiding the LLM DBA Planner toward
buffer and parallel-read controls. After DBMS-level performance
plateaus, persistent block-I/O pressure observed in block-layer and
I/O states redirects reconfiguration toward OS parameters such as
\texttt{block.read\_ahead\_kb} and \texttt{block.nr\_requests}.
With these adjustments, \textsc{AgenticDB} reduces execution time to
165.8~s, compared with AgentTune's 177.8~s.

Beyond these two mechanisms, \textsc{AgenticDB} validates each proposed
configuration before applying it by checking parameter validity,
cross-parameter compatibility, and execution requirements. If a failure still occurs, \textsc{AgenticDB} diagnoses the failure, repairs or rolls back the affected changes, and verifies that the database service has recovered (addressing C3).

We conduct extensive experiments with \textsc{AgenticDB} on MySQL and
PostgreSQL using YCSB, Sysbench, and TPC-H. Compared with SOTA methods,
\textsc{AgenticDB} achieves the best performance across all evaluated
workloads, outperforming the best-performing baseline by 118.1\% on
average and reducing the total time-to-best across workloads by
22.6\%. Further analyses show that OS-level tuning brings additional
performance improvements, validation and recovery improve
reconfiguration reliability, and consolidated experience helps
\textsc{AgenticDB} reach high-performing configurations earlier on
workloads with similar characteristics.

This paper makes the following contributions.

\begin{itemize}

\item We propose \textsc{AgenticDB}, a self-evolving agentic framework
for database workload reconfiguration. It connects an LLM DBA Planner
with the target DBMS/OS environment to jointly reconfigure DBMS knobs
and OS parameters, while providing validation and
automatic failure recovery.

\item We design two key mechanisms.
\emph{Context-grounded bottleneck diagnosis} uses workload
characteristics, configuration state, and observed runtime behavior to
diagnose the current performance bottleneck and recommend targeted
DBMS/OS reconfiguration actions. \emph{Closed-loop context evolution}
uses observed performance and runtime-state changes to update the
bottleneck diagnosis, guide subsequent decisions, terminate the loop
when performance plateaus, and consolidate experience for reuse on
workloads with similar characteristics.

\item We conduct a comprehensive evaluation of \textsc{AgenticDB} on
MySQL and PostgreSQL using representative OLTP and OLAP workloads. The
evaluation compares \textsc{AgenticDB} with SOTA methods
and analyzes the benefits of OS-level tuning, configuration validation
and failure recovery, and consolidated experience.

\end{itemize}

\section{Background and Motivation}

Recent LLM-based agents increasingly operate as systems that combine
a large language model with task context, environment-facing tools, and an
interaction loop~\cite{react,toolformer,autogen,agent_survey}.
We refer to this surrounding system as an \emph{agent harness}.
Database workload reconfiguration naturally requires such a harness
because it involves understanding the workload, interacting with the
target DBMS/OS environment, observing execution feedback, and revising
subsequent decisions. This section motivates the design of
\textsc{AgenticDB} from three perspectives.

\textbf{Harnesses augment LLM capabilities.}
A harness augments an LLM with information and operational capabilities
beyond model inference, enabling it to observe the target environment,
execute actions, collect feedback, and revise subsequent
decisions~\cite{yang2026better}. Database workload reconfiguration is particularly well suited to such
a harness-based setting because it requires workload characteristics,
deployment and runtime context, DBMS/OS operations, execution feedback,
and repeated interaction with the target environment.
\textsc{AgenticDB} therefore connects the LLM DBA Planner with these
capabilities through an agent harness and organizes reconfiguration as
an iterative diagnose--act--observe process.

\textbf{Runtime feedback as agentic observations.}
Agentic systems use runtime feedback returned by the environment as
observations to evaluate previous actions and guide subsequent
decisions~\cite{react,voyager,chen2023llm}. In database reconfiguration, observed performance and runtime-state
changes serve as such feedback by revealing how the workload responds
to each configuration change. These observations make feedback useful not only
for judging whether a configuration improves performance, but also for
guiding subsequent tuning decisions toward more promising directions.

\textbf{Memory for context evolution and experience reuse.}
In LLM-based agents, a memory module stores information from prior
interactions and provides it as context for later decisions
~\cite{generative_agents,packer2023memgpt,zhong2024memorybank,
voyager,reflexion,zhao2024expel}. In database reconfiguration,
retaining the actions and observed outcomes from earlier rounds allows
subsequent decisions to build on the evolving context rather than treat
each configuration independently. Consolidating the experience from a completed reconfiguration process
further provides prior guidance for workloads with similar
characteristics.

\section{Positioning Relative to AgentTune}
\label{sec:agenttune-comparison}

AgentTune is the closest LLM-based agent framework to \textsc{AgenticDB}.
It decomposes DBMS knob tuning into workload analysis, knob selection,
range pruning, and configuration recommendation, and then uses a
tree-based recommender to refine configurations within the selected and
pruned knob space~\cite{agenttune}. \textsc{AgenticDB} shares the use
of LLM-based DBA-style reasoning with AgentTune, but differs in how it
defines, executes, and controls the reconfiguration process.

\begin{table}[H]
\vspace{-3mm}
\centering
\small
\captionsetup{skip=2pt}
\caption{Key differences between AgentTune and \textsc{AgenticDB}.}
\label{tab:agenttune-agenticdb}

\setlength{\tabcolsep}{3pt}
\renewcommand{\arraystretch}{0.92}

\begin{tabular}{p{0.2\linewidth}p{0.31\linewidth}p{0.35\linewidth}}
\toprule
\textbf{Criterion} & \textbf{AgentTune} & \textbf{\textsc{AgenticDB}} \\
\midrule
Reconfiguration scope &
Selected DBMS knobs with pruned ranges &
Broad scope covering DBMS knobs and OS parameters \\
Feedback and iteration &
Performance-guided knob-value refinement &
Bottleneck diagnosis and closed-loop context evolution \\
Failure handling &
Candidate filtering &
Configuration validation, change verification, and automatic recovery \\
Loop control &
Fixed replay budget &
Continuation, redirection, DBMS-to-OS switching, and termination \\
\bottomrule
\end{tabular}
\vspace{-3mm}
\end{table}

Table~\ref{tab:agenttune-agenticdb} compares the two systems along four
dimensions. First, AgentTune selects DBMS knobs and prunes their ranges
before recommendation, whereas \textsc{AgenticDB} maintains a broad
reconfiguration scope covering both DBMS knobs and OS parameters.
Second, AgentTune uses performance feedback to refine knob values,
whereas \textsc{AgenticDB} combines bottleneck diagnosis with
closed-loop context evolution, allowing later rounds to focus on
unresolved bottlenecks. Third, AgentTune filters invalid candidates
before evaluation, whereas \textsc{AgenticDB} validates each proposed
configuration before applying it, verifies that the changes take effect,
and automatically recovers from failures, thereby reducing failed
applications and manual repair. Finally, AgentTune operates under a fixed replay budget, whereas
\textsc{AgenticDB} adaptively controls whether to continue, redirect,
switch from DBMS knobs to OS parameters, or terminate reconfiguration,
thereby avoiding ineffective tuning rounds when further gains are
unlikely.

\section{AgenticDB Design}
\label{sec:design}

This section presents the design of \textsc{AgenticDB}, a self-evolving
agentic framework for database workload reconfiguration. Given a target
DBMS/OS environment, a workload, an initial configuration, and a set of
tunable DBMS knobs and OS parameters, \textsc{AgenticDB} organizes
reconfiguration as a closed-loop process around
\emph{context-grounded bottleneck diagnosis} and
\emph{closed-loop context evolution}.

\begin{figure*}[t]
    \centering
    \includegraphics[width=0.85\textwidth]{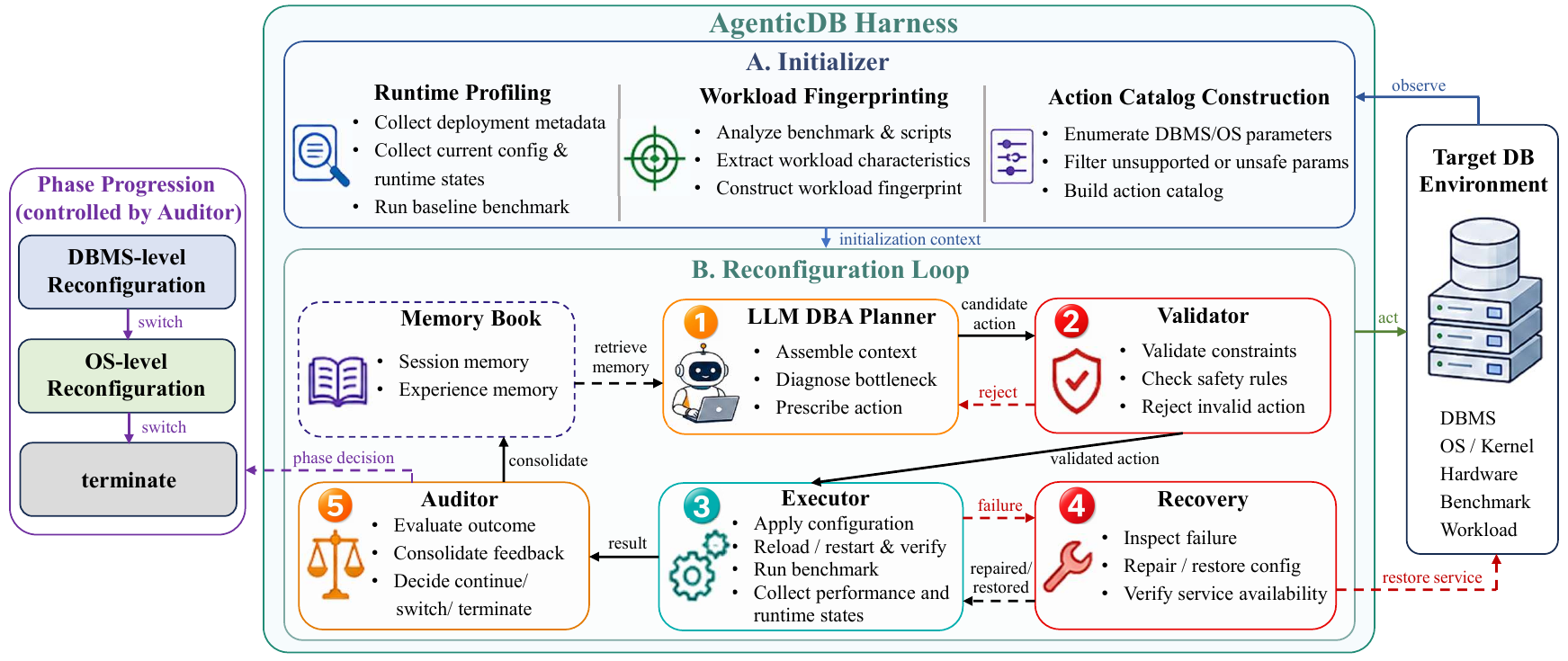}
    \captionsetup{skip=1pt}
    \caption{Overview of the \textsc{AgenticDB} framework.}
    \label{fig:agenticdb-overview}
    \vspace{-2mm}
\end{figure*}

Figure~\ref{fig:agenticdb-overview} shows how the components of
\textsc{AgenticDB} jointly realize these two mechanisms. The Initializer
profiles the workload and target environment and establishes the
reconfiguration scope covering DBMS knobs and OS parameters. Based on
the resulting initial context, the LLM DBA Planner diagnoses the current
performance bottleneck and recommends the next reconfiguration action.
The proposed configuration is then processed by the Validator,
Executor, and Recovery, which support configuration validation,
application, and failure recovery. At the end of each round, the Auditor consolidates the round outcome and determines how the loop should proceed. The Memory Book stores the consolidated outcome as context for subsequent rounds. Once the process terminates, the Auditor summarizes the accumulated actions and outcomes as reusable experience for workloads with similar characteristics.

The following subsections describe these components in the
reconfiguration process.

\subsection{Initializer}
\label{sec:initializer}

The Initializer prepares the initial context for a target
DBMS/OS environment. It profiles the environment and workload and
establishes the reconfiguration scope through three functions: runtime
profiling, workload fingerprinting, and action-catalog construction.

\textbf{Runtime profiling.}
The Initializer first records the deployment metadata, including the
DBMS and OS versions, hardware resources, current DBMS/OS
configuration, and service status. It then executes the target workload
under the initial configuration to measure the baseline performance and
collect the corresponding runtime states. For MySQL, these runtime
states come from
\texttt{SHOW GLOBAL STATUS} and
\texttt{INFORMATION\_SCHEMA.INNODB\_METRICS}. For PostgreSQL, they come
from
\texttt{pg\_stat\_database},
\texttt{pg\_stat\_activity},
\texttt{pg\_locks},
\texttt{pg\_stat\_bgwriter},
\texttt{pg\_stat\_wal}, and
\texttt{pg\_stat\_io}. The collected runtime states cover buffer
behavior, I/O and logging activity, connection and lock status, and
other execution behavior.

\textbf{Workload fingerprinting.}
The Initializer constructs a structured workload fingerprint that
summarizes the current workload and DBMS environment. As shown in
Table~\ref{tab:fingerprint}, the fingerprint records the DBMS version,
benchmark identity, workload class, base workload type, access patterns, workload scale, and optimization objective.

\begin{table}[H]
\vspace{-2mm}
\centering
\captionsetup{skip=2pt}
\caption{Workload fingerprint fields used by \textsc{AgenticDB}.}
\label{tab:fingerprint}

\begingroup
\small
\setlength{\tabcolsep}{3pt}
\renewcommand{\arraystretch}{0.92}
\setlength{\aboverulesep}{0.25ex}
\setlength{\belowrulesep}{0.25ex}

\begin{tabularx}{\columnwidth}{
    @{}
    >{\raggedright\arraybackslash}p{0.20\columnwidth}
    >{\raggedright\arraybackslash}p{0.33\columnwidth}
    >{\raggedright\arraybackslash}X
    @{}
}
\toprule
\textbf{Field}
& \textbf{Purpose}
& \textbf{Representative values} \\
\midrule

DBMS version
& Identify the DBMS family and version for compatible experience retrieval.
& \texttt{mysql-8.0}, \texttt{postgresql-16}, etc. \\

Benchmark identity
& Identify the workload source and benchmark script.
& YCSB, Sysbench, TPC-H, custom workload, etc. \\

Workload class
& Capture the high-level workload family.
& \texttt{oltp}, \texttt{olap}, \texttt{mixed}. \\

Base workload type
& Capture coarse read/write tendency.
& \texttt{read\_only}, \texttt{write\_only}, \texttt{read\_write}, \texttt{analytical}, etc. \\

Access patterns
& Capture dominant access and operator behavior.
& Point lookup, range scan, index scan, sequential scan, join-heavy, aggregation/sort, etc. \\

Workload scale
& Capture the workload intensity and data scale.
& Client/thread count, table size, data volume, query count, request rate, etc. \\

Optimization objective
& Capture the performance goal of the reconfiguration task.
& Throughput-sensitive, latency-sensitive, minimize execution time, etc. \\

\bottomrule
\end{tabularx}
\endgroup
\vspace{-2mm}
\end{table}

The workload fingerprint provides the LLM DBA Planner with a structured
description of the workload characteristics and optimization objective
for the initial planning round. It also serves as the retrieval key for
experience memory, enabling relevant experience to be retrieved from
workloads with similar characteristics.

\textbf{Action-catalog construction.}
The Initializer builds an action catalog that defines a broad
reconfiguration scope covering DBMS knobs and OS parameters. For the
DBMS layer, the catalog retains nearly all tunable knobs exposed by the
target environment, excluding only those that are read-only,
unsupported, deprecated, or explicitly unsafe. For the OS layer, it
covers parameters related to memory management, dirty-page writeback,
I/O scheduling, CPU scheduling, file-system behavior, and kernel
resource limits.

\subsection{LLM DBA Planner}
\label{sec:planner}

The LLM DBA Planner is the decision-making core of
\textsc{AgenticDB} and implements the first mechanism:
\emph{context-grounded bottleneck diagnosis}. In the initial round, it
uses the initialization context and, when available, relevant experience
retrieved from workloads with similar characteristics to diagnose the
current performance bottleneck and recommend the first reconfiguration
action. In subsequent rounds, it additionally incorporates information
from previous rounds to update the diagnosis and recommend the next reconfiguration action.

\begin{figure*}[t]
    \centering
    \includegraphics[width=0.8\textwidth]{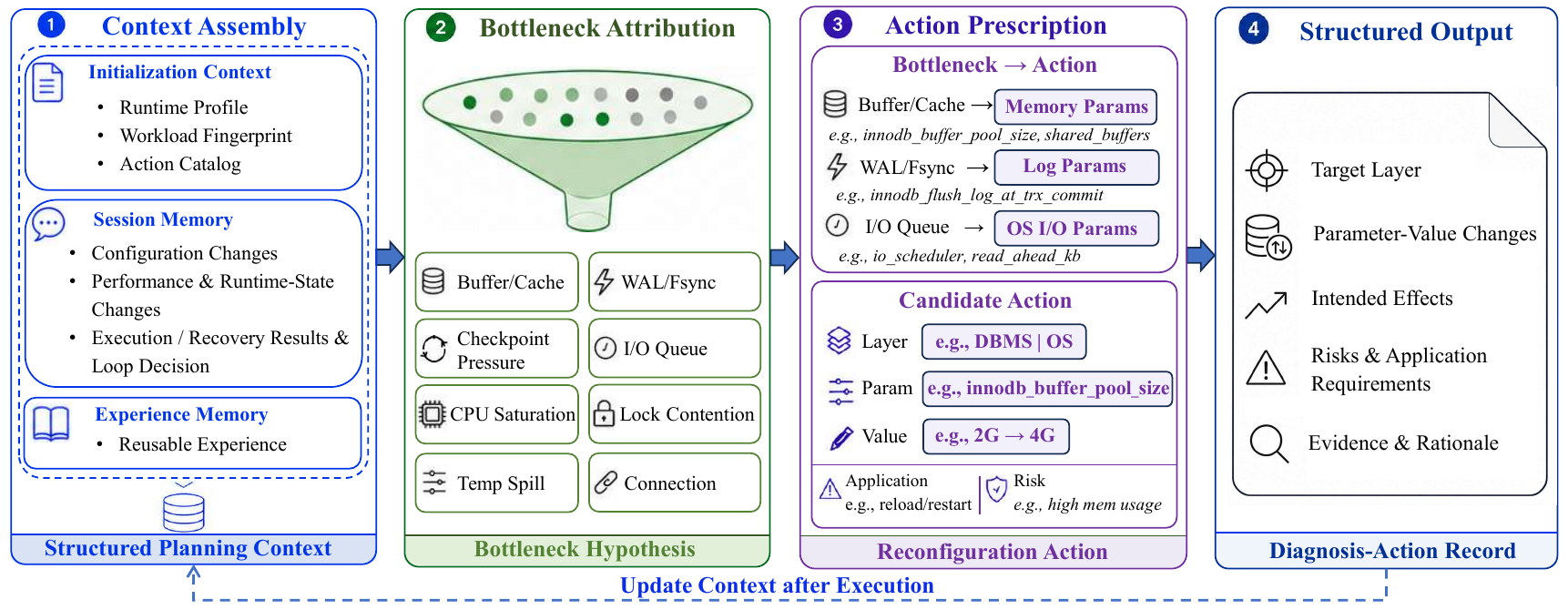}
    \captionsetup{skip=2pt}
    \caption{Context-grounded bottleneck diagnosis and configuration action prescription in the LLM DBA Planner.}
    \label{fig:planner}
    \vspace{-2mm}
\end{figure*}

Figure~\ref{fig:planner} illustrates the workflow of the LLM DBA
Planner, which consists of four steps: context assembly, bottleneck
attribution, action prescription, and structured output.

\textbf{Context assembly.}
At each planning round, the LLM DBA Planner assembles the planning
context from three sources: the initialization context, session memory,
and experience memory.

The \emph{initialization context} is produced by the Initializer and
consists of the runtime profile, workload fingerprint, and action
catalog. Together, they describe the deployment metadata, baseline
performance and runtime states, workload characteristics and
optimization objective, and the reconfiguration scope covering DBMS
knobs and OS parameters.

The \emph{session memory} stores the outcomes consolidated by the
Auditor from previous rounds of the current reconfiguration process.
These outcomes include the applied configuration changes, observed
performance and runtime-state changes, execution and recovery results,
and the decision on how the loop should proceed.

The \emph{experience memory} contains reusable summaries from completed
reconfiguration processes. It is retrieved using the current workload
fingerprint and provides guidance based on workloads with similar
characteristics, including useful reconfiguration directions,
ineffective or failed changes, and recurring bottlenecks.

Together, these sources allow the LLM DBA Planner to reason with the
initial state, the outcomes of the current process, and relevant
experience from previous workloads.

\textbf{Bottleneck attribution.}
Given the planning context, the LLM DBA Planner treats the observed
runtime behavior and performance changes as diagnostic evidence for
bottleneck attribution. Figure~\ref{fig:planner} illustrates the
representative bottleneck hypotheses used in this process.
Buffer/cache pressure indicates insufficient data caching or frequent
physical reads. WAL/fsync pressure reflects transaction commit overhead
caused by frequent log flushes or synchronization. Checkpoint pressure
reflects dirty-page accumulation and background flushing overhead.
I/O queue pressure indicates queuing along the storage path or high
device latency. CPU saturation indicates compute-bound execution, while
lock contention reflects blocking among concurrent transactions.
Temporary spill indicates that sort, join, or aggregation operations
spill intermediate data to disk. Connection pressure reflects limits or
overhead in session, thread, or connection management.

The LLM DBA Planner weighs the available diagnostic evidence to identify
the current performance bottleneck. For example, a write-heavy workload
with high commit latency and intensive WAL activity may indicate
WAL/fsync pressure, whereas a read-intensive workload with frequent
physical reads and poor buffer behavior may indicate buffer/cache
pressure. Once the current bottleneck is identified, the LLM DBA Planner recommends configuration changes intended to alleviate it.

\textbf{Action prescription.}
After identifying the current bottleneck, the LLM DBA Planner selects
relevant parameters from the action catalog and determines their values
according to the diagnosed bottleneck and current planning context.
As illustrated in Figure~\ref{fig:planner}, buffer/cache pressure may
lead to memory-related parameters such as
\texttt{innodb\_buffer\_pool\_size} and
\texttt{shared\_buffers}, while WAL/fsync pressure may lead to logging
parameters such as
\texttt{innodb\_flush\_log\_at\_trx\_commit}. At the OS level, I/O
queue pressure may lead to parameters such as
\path|io_scheduler| and \path|read_ahead_kb|. For each proposed
change, it explains how the parameter is expected to alleviate the
diagnosed bottleneck.

The reconfiguration process starts with DBMS knobs. In the initial
round, the LLM DBA Planner proposes a broader DBMS configuration
covering multiple knobs relevant to the workload and initial diagnosis.
In later rounds, it makes more focused changes according to the observed
outcomes: beneficial changes may be retained and further adjusted,
ineffective changes may be revised or removed, and changes associated
with validation or execution failures are avoided. Once DBMS-level
performance plateaus, reconfiguration shifts from DBMS knobs to OS
parameters.

\textbf{Structured output.}
The LLM DBA Planner emits each recommendation as a structured
diagnosis--action record. As shown in Figure~\ref{fig:planner}, the
record contains the diagnosed bottleneck, target layer, relevant
parameter--value changes, intended effects, potential risks, application
requirements, and supporting evidence and rationale. The application
requirements indicate whether the proposed changes can be applied
dynamically or require a reload or restart. This record provides
structured input for subsequent validation, execution, and auditing.

\subsection{Validator}
\label{sec:validator}

The Validator checks each proposed configuration before it is applied
to the target DBMS/OS environment. Although the LLM DBA Planner grounds
its recommendations in the planning context, LLM-generated outputs may
still hallucinate nonexistent parameters or misinterpret parameter
semantics, deployment constraints, and application requirements. The
Validator therefore checks both individual parameter--value changes and
the proposed configuration as a whole before execution.

The Validator does not rely on LLM reasoning. Instead, it applies
predefined checks using parameter metadata, the current configuration,
hardware limits, benchmark rules, and known safety constraints. It first
checks whether the proposed action is permitted at the current DBMS- or
OS-level stage and whether its action type is supported. It then verifies parameter existence, normalizes parameter value types
and units, and checks enum values, allowed ranges, and hardware limits. The Validator also rejects changes that modify protected benchmark
settings or benchmark semantics, involve known dangerous parameters, or
trigger known parameter conflicts. Finally, it verifies the application requirements specified by the LLM
DBA Planner against runtime parameter metadata and rejects actions with inconsistent requirements.

The Validator returns the validation result, normalized configuration,
verified application requirements, and any failed checks. Rejected
proposals are returned to the LLM DBA Planner for revision, while only
validated configurations are passed to the Executor for application and
workload evaluation.

\subsection{Executor}
\label{sec:executor}

The Executor applies each validated configuration to the target
DBMS/OS environment. Based on the verified application requirements,
it uses the corresponding DBMS or OS interface to apply the changes
dynamically or through a reload or restart. After application, the
Executor verifies that the changes have taken effect. Once verification
succeeds, it runs the benchmark workload and collects the resulting
performance and runtime states. The collected results are then passed
to the Auditor for evaluation and consolidation.

If configuration application, reload or restart, effective-value
verification, or workload execution fails, the Executor aborts the
current execution attempt and passes the failure status, relevant
logs, and configuration changes attempted in the round to Recovery.

\subsection{Recovery}
\label{sec:recovery}

The Recovery component handles execution failures reported by the
Executor. Although each proposed configuration has passed the
Validator's checks, some failures become observable only during
configuration application, DBMS reload or restart, or workload
execution under the updated configuration. Such failures may result
from restart-time incompatibilities, unexpected parameter interactions,
or deployment-specific conditions that cannot be fully verified before
execution.

When a failure occurs, Recovery examines the service status, DBMS and
system logs, effective configuration values, and the configuration
changes attempted in the current round. It repairs the faulty
configuration when possible; otherwise, it restores the last safe
DBMS/OS configuration and performs the required reload, restart, or OS
parameter rollback. Recovery then verifies that the database service is
available and returns the repaired or restored state to the Executor.
The recovery result is included in the round outcome passed to the
Auditor, allowing subsequent rounds to avoid changes associated with
similar failures.

\subsection{Auditor}
\label{sec:auditor}

The Auditor is invoked at the end of each reconfiguration round and
provides the feedback-consolidation and loop-control functions of the
second mechanism: \emph{closed-loop context evolution}. Configuration
application and workload execution in each round produce heterogeneous
information, including applied configuration changes, benchmark
results, runtime-state changes, effective parameter values, execution
results, and any recovery results. Directly passing all raw outputs to
subsequent rounds would cause context explosion, making the planning
context increasingly long, repetitive, and noisy. The Auditor therefore
preserves the information needed for subsequent planning, discards
redundant details, and organizes the retained information into a
structured round outcome.
Table~\ref{tab:auditor-consolidation} summarizes how the Auditor
consolidates these heterogeneous round-level outputs for subsequent
planning.

\begin{table}[H]
\vspace{-2mm}
\centering
\small
\captionsetup{skip=2pt}
\caption{Round information consolidated by the Auditor.}
\label{tab:auditor-consolidation}
\begin{tabular}{p{0.23\linewidth}p{0.32\linewidth}p{0.36\linewidth}}
\toprule
\textbf{Session-memory item}
& \textbf{Consolidated from}
& \textbf{Role in subsequent rounds} \\
\midrule

Applied configuration changes
& Proposed and effective parameter values.
& Records the changes applied in the current round. \\

Performance and runtime-state changes
& Benchmark results and runtime states observed during execution.
& Describes the performance impact and updated runtime behavior. \\

Execution and recovery results
& Application status, failure logs, and recovery outcome.
& Identifies failed or risky changes and the recovered state. \\

Loop decision
& Current round outcome and previous loop decisions.
& Records whether to continue, redirect, switch, or terminate. \\

\bottomrule
\end{tabular}
\vspace{-2mm}
\end{table}

Based on the consolidated round outcome, the Auditor controls how the
reconfiguration process proceeds. During DBMS-level reconfiguration, it
continues the current direction while performance gains remain,
redirects reconfiguration when further gains diminish or the
bottleneck changes, and switches to OS-level reconfiguration when
DBMS-level performance plateaus. During OS-level reconfiguration, it
continues while additional gains remain and terminates when no
substantial improvement remains. These decisions are made within a user-specified maximum number of
reconfiguration rounds, which bounds the overall reconfiguration cost.

The consolidated round outcome is stored in session memory as context
for subsequent rounds. Once the process terminates,
the Auditor summarizes the accumulated actions and outcomes as reusable
experience for workloads with similar characteristics.

\subsection{Memory Book}
\label{sec:memory-book}

The Memory Book works with the Auditor to support the second mechanism: \emph{closed-loop context evolution}, by storing the information
accumulated during reconfiguration. In \textsc{AgenticDB},
self-evolution is reflected in the progressive update of the planning
context with observed round outcomes and reusable experience, while the
LLM parameters remain fixed. The Memory Book stores these two types of
information as session memory and experience memory, respectively,
which the LLM DBA Planner accesses when assembling the planning context
for each round.

\textbf{Session memory.}
Session memory stores the consolidated round outcomes from the current
reconfiguration process. At the end of each round, the Auditor
consolidates the applied configuration changes, observed performance
and runtime-state changes, execution and recovery results, and loop
decision. The Memory Book stores this information in session memory.
When assembling the planning context for the next round, the LLM DBA
Planner retrieves relevant entries from session memory. Subsequent
decisions can therefore build on changes that improved performance,
bottlenecks that remain unresolved, and actions that were ineffective
or caused failures.

\textbf{Experience memory.}
Experience memory stores reusable summaries from completed
reconfiguration processes. Once a process terminates, the Auditor
summarizes the accumulated actions and outcomes, and the Memory Book
stores the resulting experience using the workload fingerprint as the
retrieval key. The stored experience records recurring bottlenecks,
beneficial reconfiguration directions, and ineffective or failed
changes. For a new process, the LLM DBA Planner retrieves relevant
experience from workloads with similar characteristics and incorporates
it into the initial planning context.

\begin{figure}[H]
\vspace{-2mm}
    \centering
    \includegraphics[width=1\columnwidth]{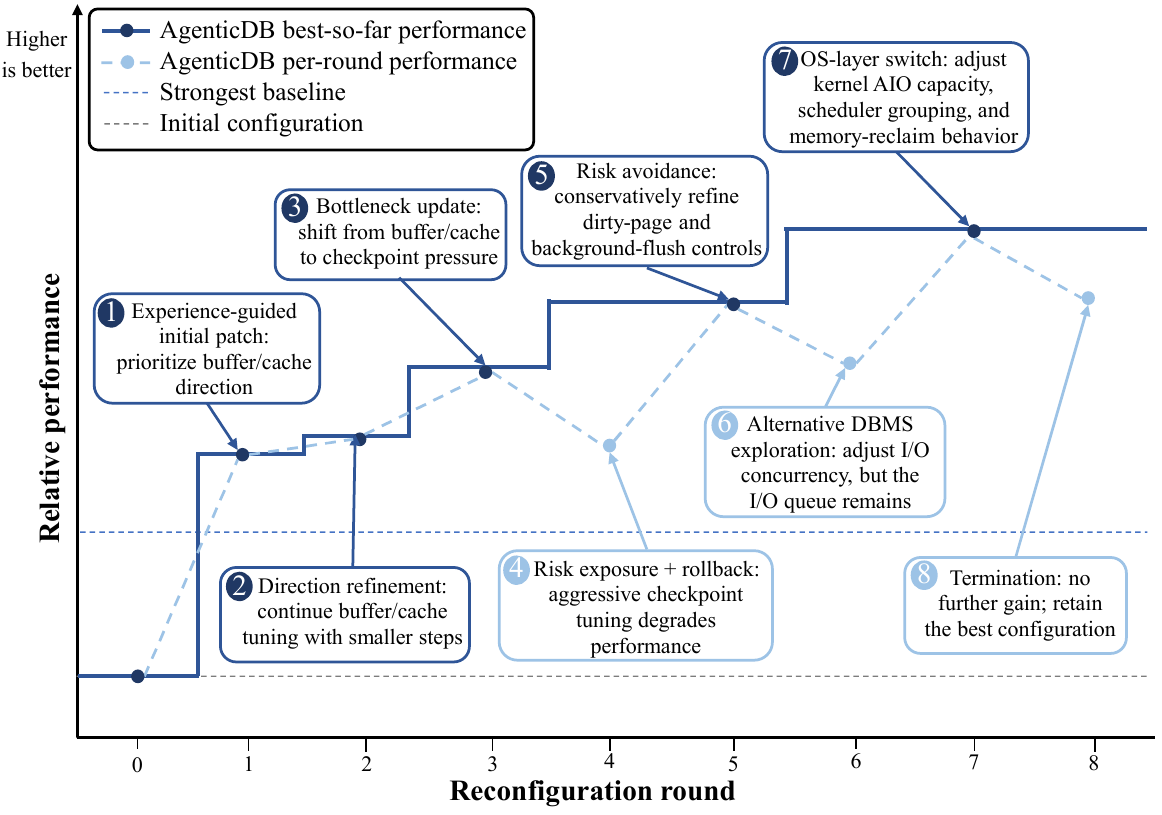}
    \captionsetup{skip=2pt}
    \caption{Illustrative example of closed-loop context evolution
    across reconfiguration rounds.}
    \label{fig:memory-evolution}
    \vspace{-2mm}
\end{figure}

\textbf{Closed-loop context evolution.}
Figure~\ref{fig:memory-evolution} illustrates the closed-loop context
evolution process driven by the Memory Book and the Auditor, in which
the planning context progressively evolves across reconfiguration
rounds and processes.

Starting from the initial configuration, Rounds~1--3 use the
initialization context, relevant experience, and context accumulated in
session memory to move from a broad buffer/cache-oriented DBMS
configuration to more focused adjustments, and then redirect the
diagnosis toward checkpoint pressure. Round~4
explores a more aggressive checkpoint configuration, but the resulting
performance degradation shows that the change is ineffective. Recovery
restores the previous safe configuration, and the failure is recorded
in session memory. Rounds~5--6 avoid repeating this change, refine
dirty-page and background-flush controls, and test DBMS-level I/O
concurrency. The remaining I/O pressure and diminishing performance
gains indicate that DBMS-level performance has plateaued. Round~7
therefore shifts to OS-level reconfiguration and improves performance
through AIO, CPU-scheduling, and memory-reclamation adjustments.
Round~8 produces no further gain, so the Auditor terminates the process
with the best stable configuration found.

Across these rounds, the Auditor compares the expected and realized
outcomes, consolidates the evidence needed for later decisions, and
determines how the loop should proceed. The Memory Book carries this
information into subsequent rounds through session memory and retains
the final summary through experience memory. This allows the LLM DBA
Planner to progressively refine its bottleneck diagnosis and parameter
choices, preserve beneficial changes, avoid ineffective or failed
changes, and shift from DBMS knobs to OS parameters when DBMS-level
performance plateaus. Thus, each round builds on previously accumulated
evidence rather than acting as an independent trial, realizing the
self-evolving reconfiguration process of \textsc{AgenticDB} without
updating the LLM parameters.

\section{Evaluation Methodology}
\label{sec:evaluation-methodology}

\subsection{Research Questions}
\label{sec:rq}

We evaluate \textsc{AgenticDB} through the following research questions:

\begin{itemize}
    \item \textbf{RQ1:} How does \textsc{AgenticDB} compare with
    representative database tuning baselines in performance and tuning
    efficiency across different workloads and DBMSs?

    \item \textbf{RQ2:} How much additional performance improvement
    does OS-level reconfiguration provide after DBMS-level performance plateaus?

    \item \textbf{RQ3:} How do the Validator and Recovery improve the reliability of the reconfiguration process?

    \item \textbf{RQ4:} How do different LLM backends and experience
memory affect the reconfiguration performance and behavior of
\textsc{AgenticDB}?
\end{itemize}

RQ1 and RQ2 evaluate the performance and tuning efficiency of
\textsc{AgenticDB} and the additional improvement provided by OS-level
reconfiguration, respectively. RQ3 examines how the Validator rejects invalid configurations before
execution and how Recovery handles execution failures.
RQ4 analyzes
performance differences among LLM backends and how experience memory
changes the early reconfiguration process.

\subsection{Experimental Setup}
\label{sec:setup}

\textbf{Testbed and DBMSs.}
All experiments are conducted on a CPU-only Linux x86\_64 server with
12~vCPUs, 64~GB of memory, and a 120~GB SSD. We evaluate MySQL~8.0.45
and PostgreSQL~16.13.

\textbf{Workloads.}
For MySQL, we use five workloads covering OLTP and OLAP: YCSB, TPC-H,
and three Sysbench workloads (\texttt{read}, \texttt{write}, and
\texttt{readwrite})~\cite{ycsb,tpch,sysbench}. YCSB represents
key-value OLTP workloads, TPC-H represents analytical queries, and the
Sysbench workloads represent read-only, write-only, and mixed
read-write OLTP behaviors, respectively.

For PostgreSQL, we use the same three Sysbench workloads to evaluate
generalization across DBMSs.

\textbf{Baselines.}
We compare \textsc{AgenticDB} with four representative database tuning
baselines.

\begin{itemize}

    \item \textbf{GPTuner}~\cite{gptuner} represents LLM-assisted
    Bayesian optimization. It uses LLM-structured tuning knowledge to
    guide knob selection, search-space reduction, and coarse-to-fine
    configuration search.

    \item \textbf{DB-BERT}~\cite{dbbert} uses a BERT-based model to
    extract tuning hints from DBMS manuals and combines the extracted
    knowledge with reinforcement learning.

    \item \textbf{AgentTune}~\cite{agenttune} is an LLM-based database
    tuning baseline. It decomposes DBMS knob tuning into workload
    analysis, knob selection, range pruning, and configuration
    recommendation, and uses a tree-based recommender to refine
    configurations within the selected knob space.

    \item \textbf{CDBTune}~\cite{cdbtune} applies deep reinforcement
    learning to iteratively optimize database configurations using
    feedback collected during tuning.

\end{itemize}

% For fairness, we use the hyperparameter settings reported in the
% original papers or released implementations.

\textbf{Metrics.}
Following common practice in database configuration tuning, we evaluate
OLTP workloads using throughput and tail latency~\cite{cdbtune,agenttune}.
For Sysbench and YCSB, throughput is measured by TPS and tail latency is
measured by P95 latency. We use $\mathrm{TPS}/\mathrm{P95}$ as the
optimization objective, where higher values indicate higher throughput
and lower tail latency.

For the OLAP workload TPC-H, we use total query execution time as the
optimization objective, where lower values are better.

\textbf{Evaluation Protocol.}
For fair comparison, all methods start from the same initial DBMS/OS
state and use the same benchmark scripts and optimization objective.
They are given a maximum wall-clock tuning budget of 100 minutes, which
serves as an upper bound; within this budget, each method follows its
original termination condition. Baseline hyperparameters follow the
settings reported in the original papers or released implementations.

Except for the LLM-backend study in RQ4, all generative-LLM-based
methods, including \textsc{AgenticDB}, GPTuner, and AgentTune, use
GPT-5.5~\cite{openai-gpt55} as the underlying LLM backend. For YCSB and Sysbench, each benchmark run lasts 60 seconds; for TPC-H,
each run executes the complete TPC-H query set. To account for randomness
in both reconfiguration and benchmark execution, we repeat each method
three times for every workload and report the run with the median final
objective value.

\textbf{Memory Protocol.}
All experiments retain session memory during each reconfiguration
process. Experience memory is enabled only in the dedicated memory
study in RQ4; all other experiments disable it. In this study,
experience memory is constructed from reusable summaries of completed
reconfiguration processes with similar workload characteristics. To
prevent leakage, the experience memory used for a target evaluation
does not include execution outcomes or final results produced during
that target evaluation. The retrieved experience therefore provides
prior guidance from other completed processes rather than access to the
target evaluation outcome.

\section{Experimental Results}
\label{sec:results}

In this section, we answer the research questions introduced in Section~\ref{sec:rq}.

\begin{figure*}[t]
    \centering

    \includegraphics[width=0.73\textwidth]{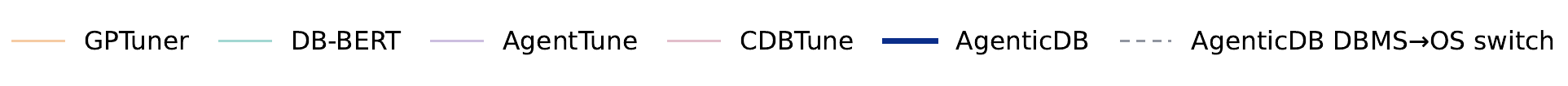}

    \vspace{-1mm}

    \begin{tabular}{c@{\hspace{5pt}}c@{\hspace{5pt}}c}
    % Left column
    \begin{minipage}{0.465\textwidth}
        \centering

        % PG Read
        \includegraphics[width=0.32\linewidth]{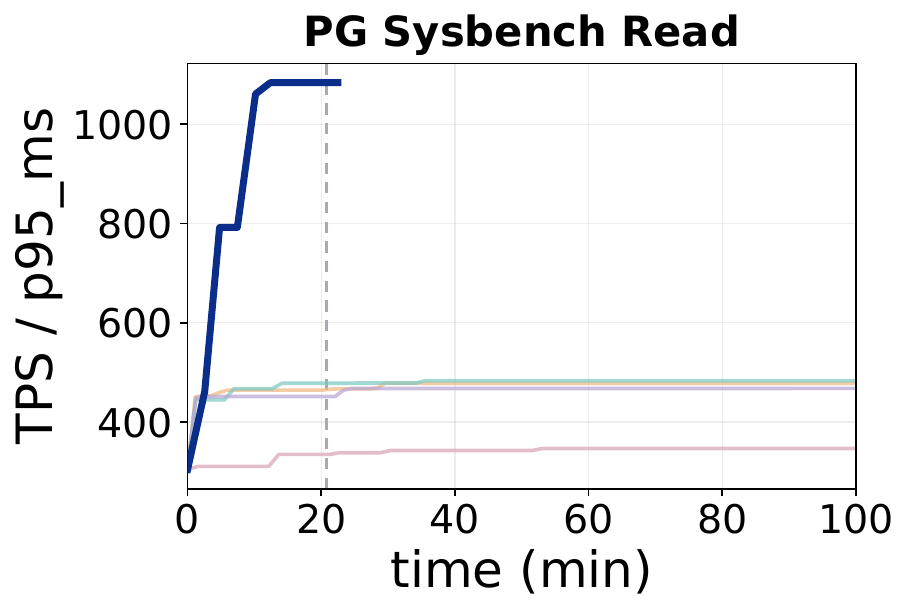}
        \includegraphics[width=0.32\linewidth]{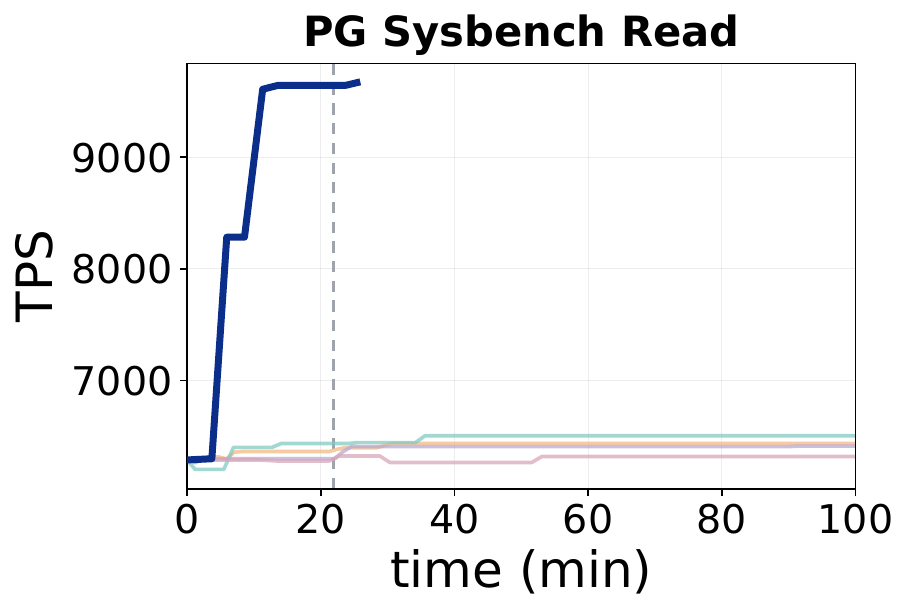}
        \includegraphics[width=0.32\linewidth]{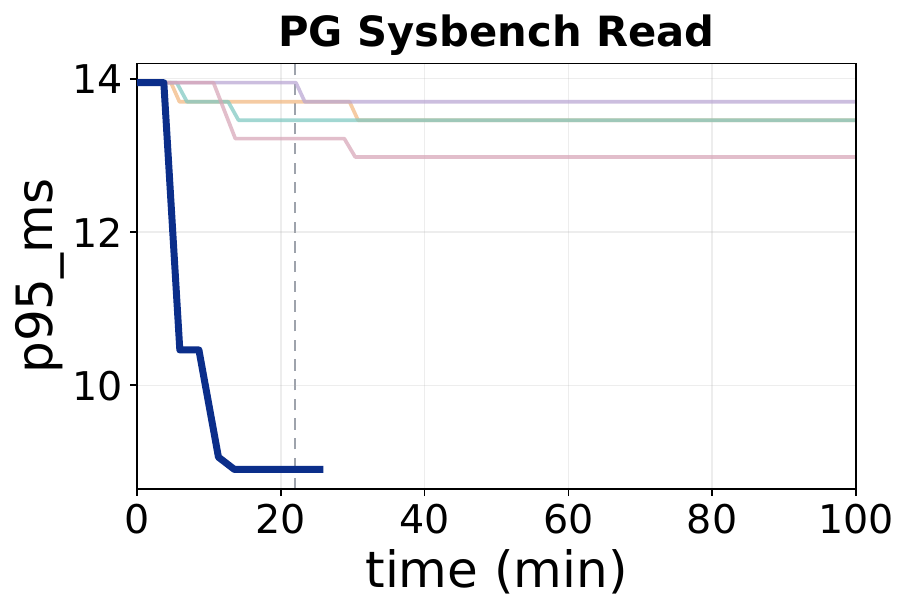}

        \vspace{1mm}

        % PG ReadWrite
        \includegraphics[width=0.32\linewidth]{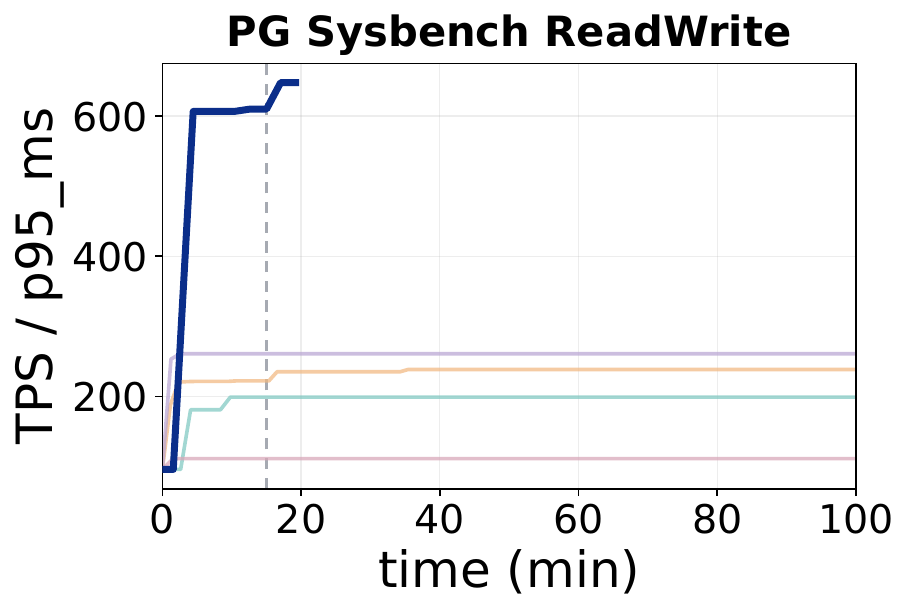}
        \includegraphics[width=0.32\linewidth]{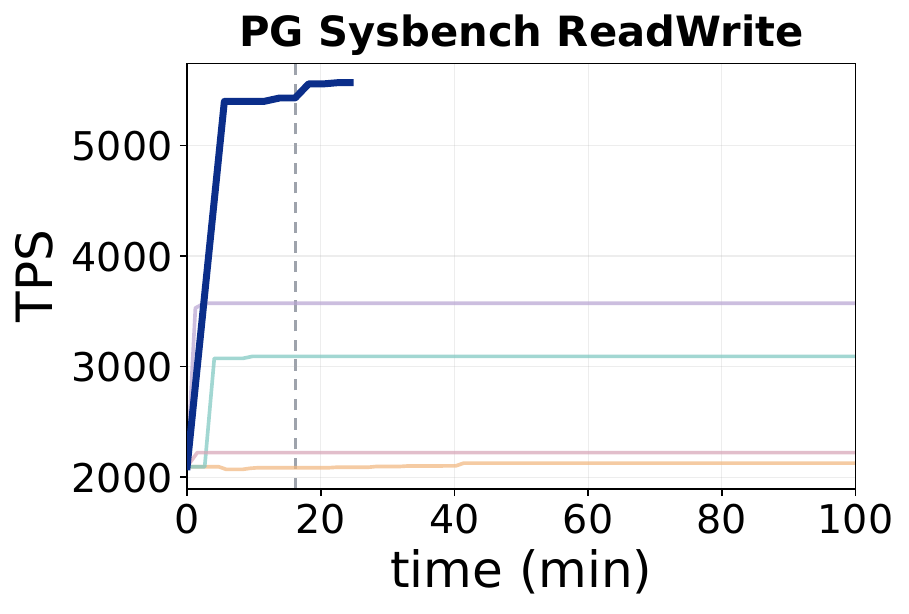}
        \includegraphics[width=0.32\linewidth]{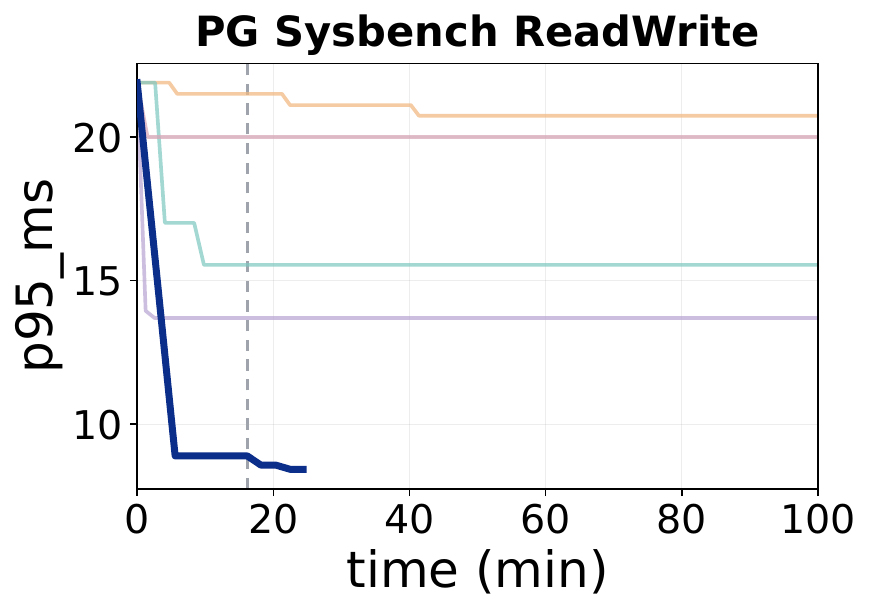}

        \vspace{1mm}

        % MS Read
        \includegraphics[width=0.32\linewidth]{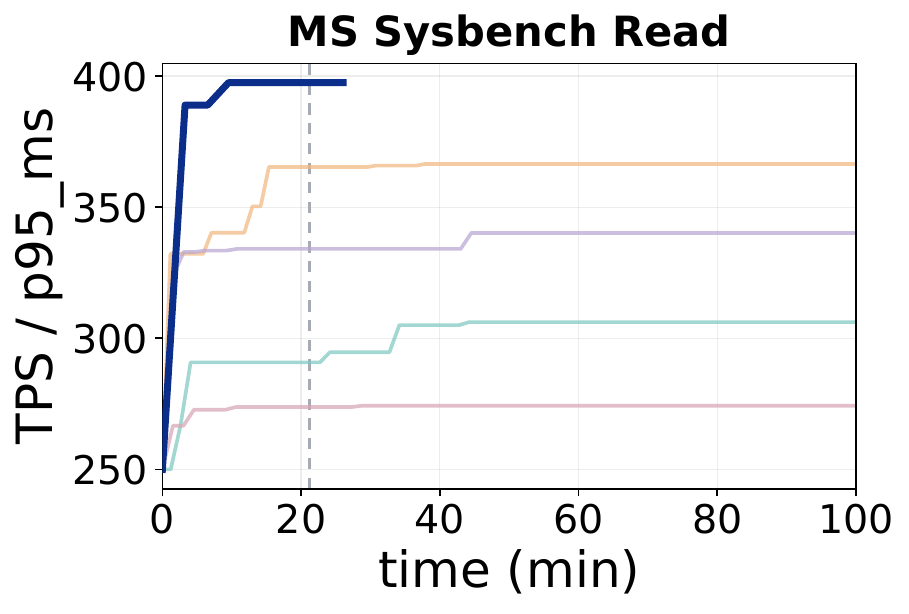}
        \includegraphics[width=0.32\linewidth]{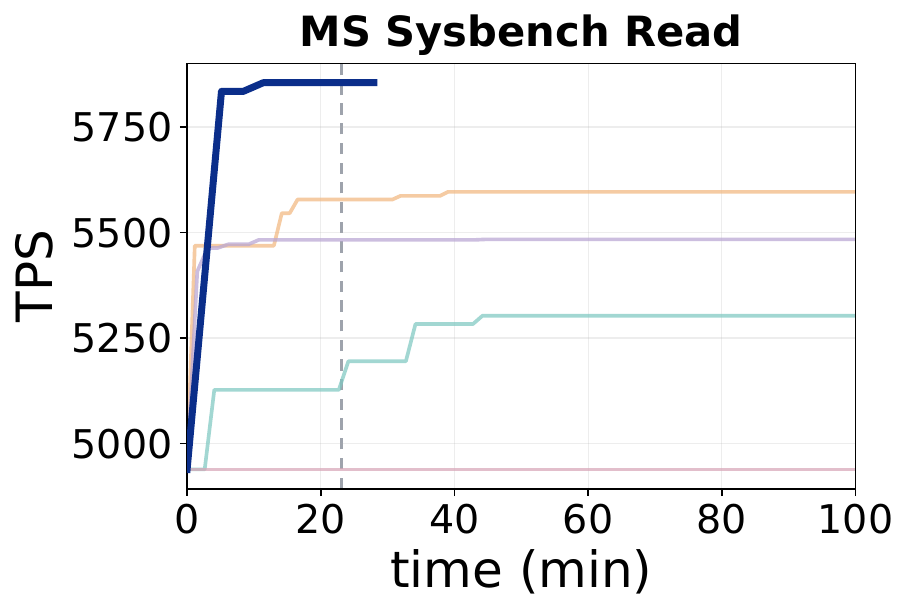}
        \includegraphics[width=0.32\linewidth]{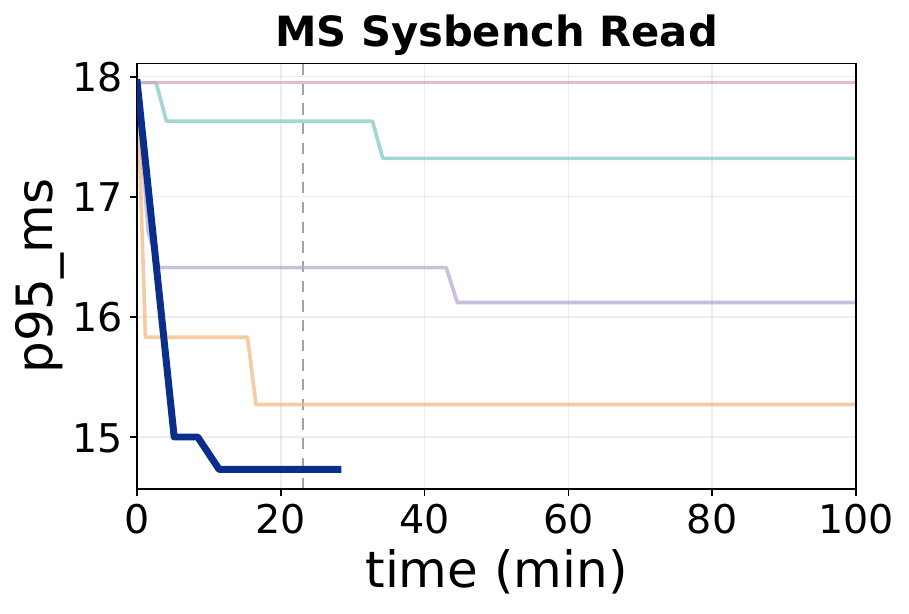}

        \vspace{1mm}

        % MS ReadWrite
        \includegraphics[width=0.32\linewidth]{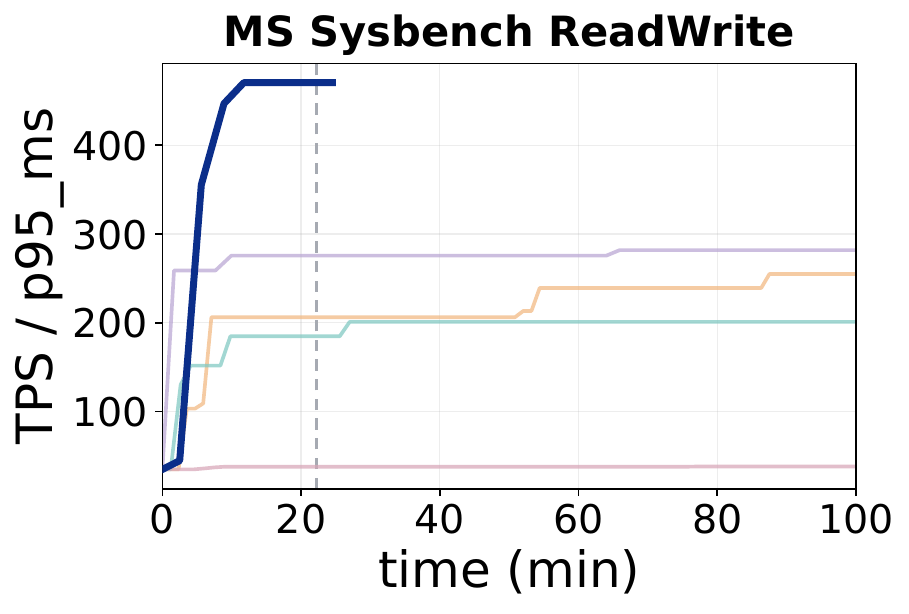}
        \includegraphics[width=0.32\linewidth]{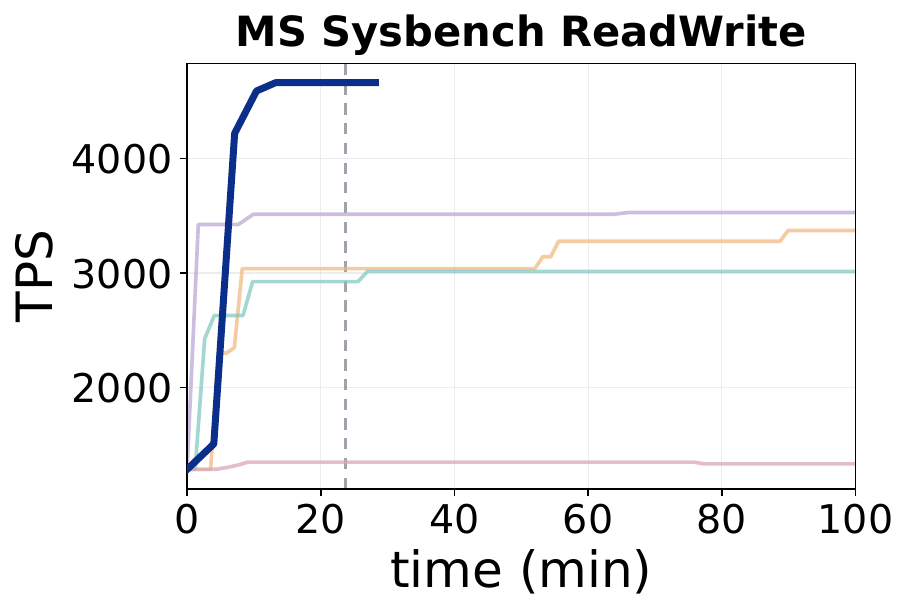}
        \includegraphics[width=0.32\linewidth]{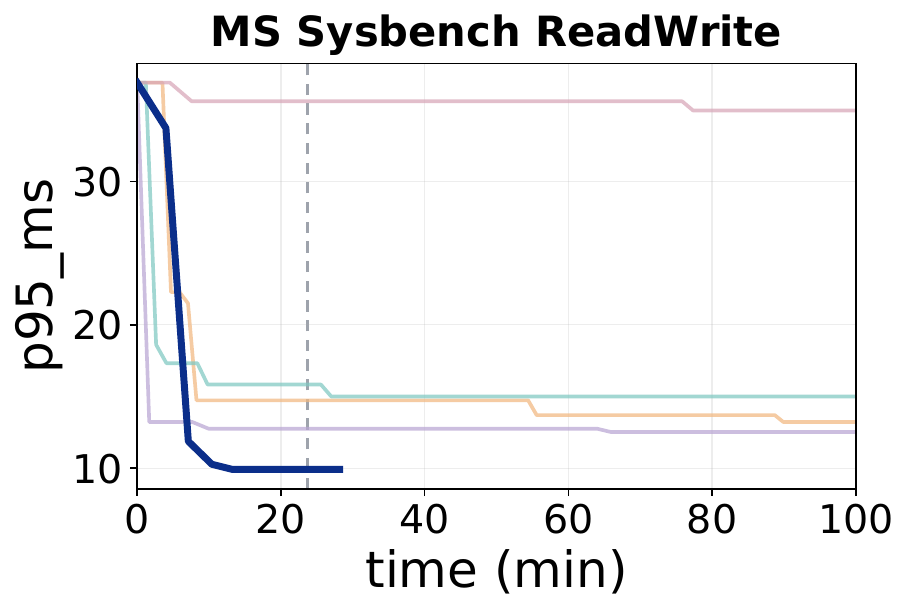}
    \end{minipage}
    &
    % Middle dashed line
    \raisebox{-0.49\height}{
    \begin{tikzpicture}
        \draw[densely dashed, gray, line width=0.35pt] (0,0) -- (0,7.3cm);
    \end{tikzpicture}
    }
    &
    % Right column
    \begin{minipage}{0.465\textwidth}
        \centering

        % PG Write
        \includegraphics[width=0.32\linewidth]{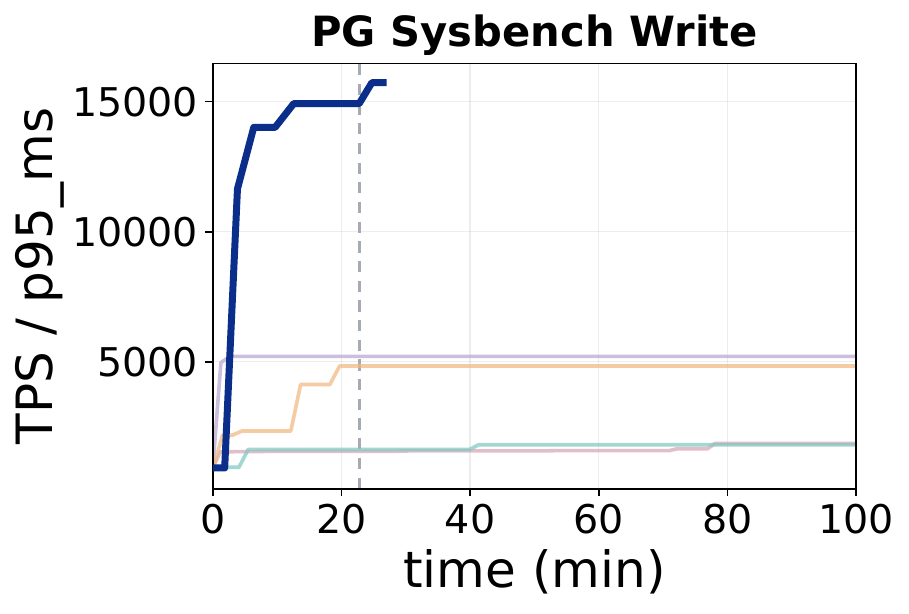}
        \includegraphics[width=0.32\linewidth]{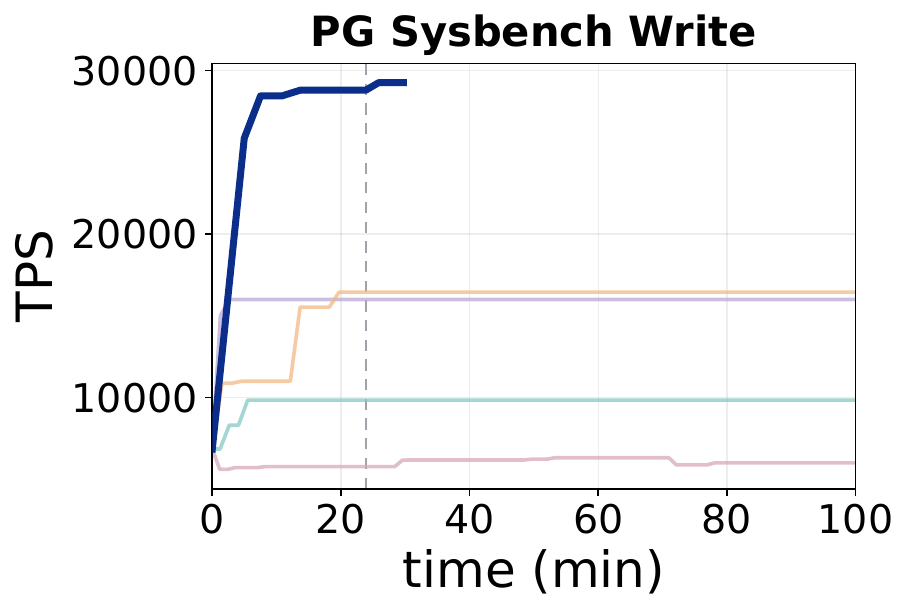}
        \includegraphics[width=0.32\linewidth]{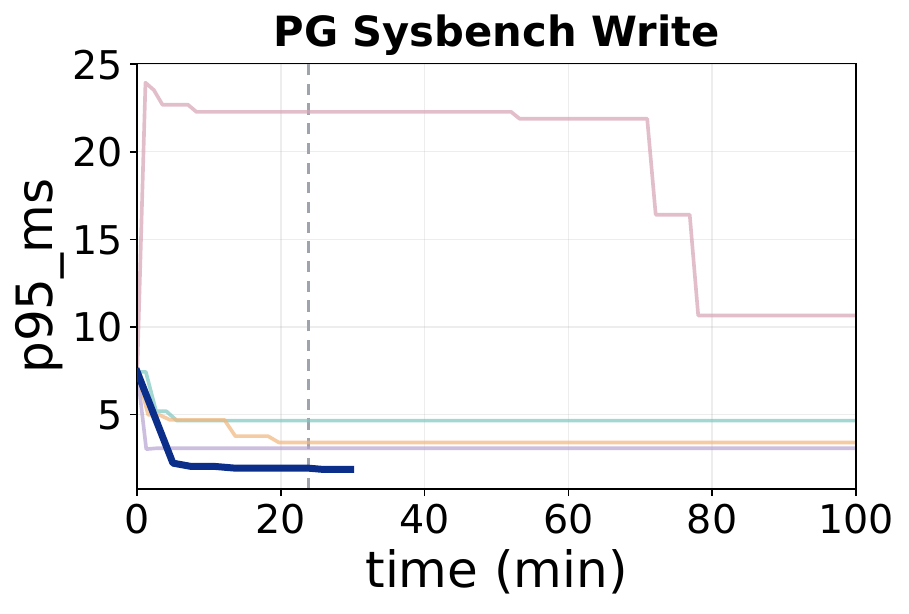}

        \vspace{1mm}

        % MS YCSB
        \includegraphics[width=0.32\linewidth]{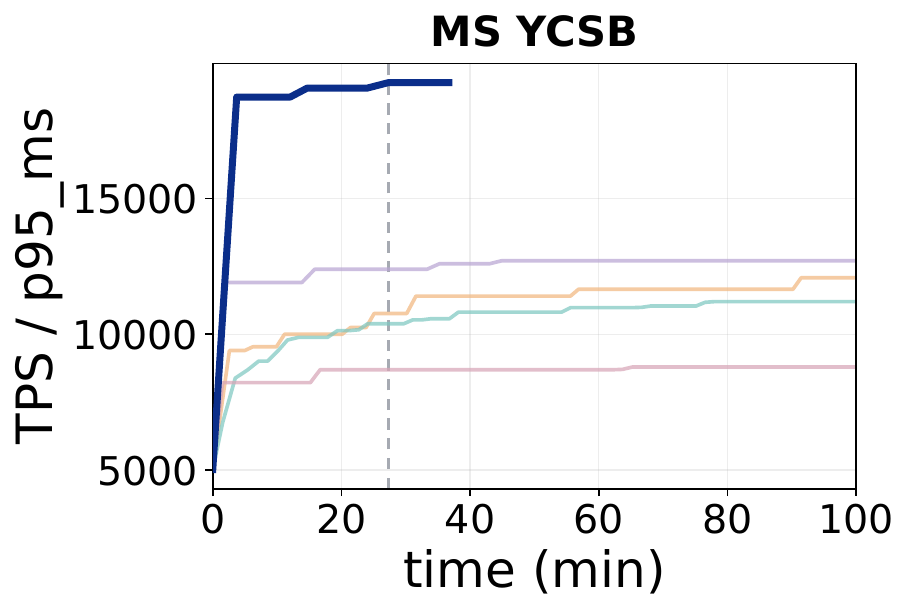}
        \includegraphics[width=0.32\linewidth]{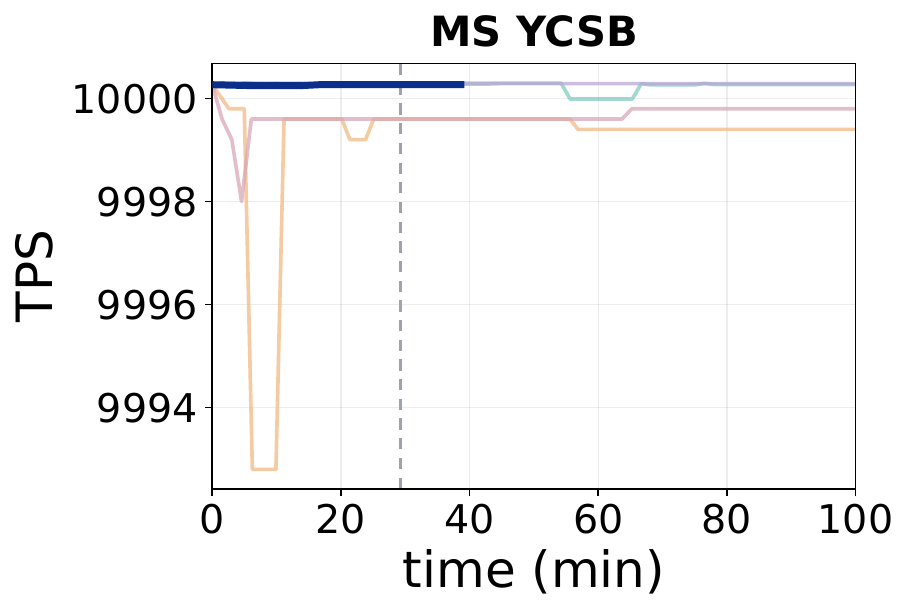}
        \includegraphics[width=0.32\linewidth]{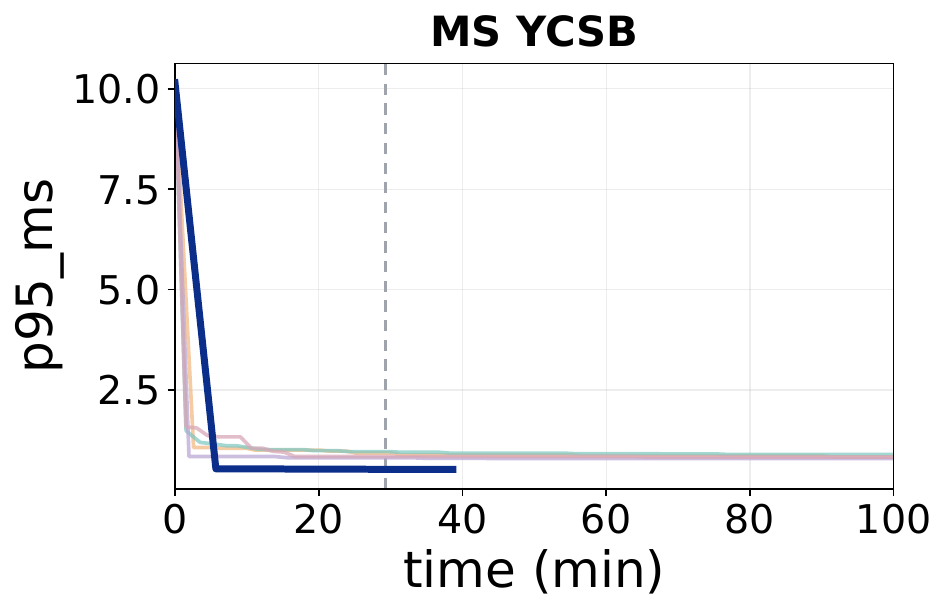}

        \vspace{1mm}

        % MS Write
        \includegraphics[width=0.32\linewidth]{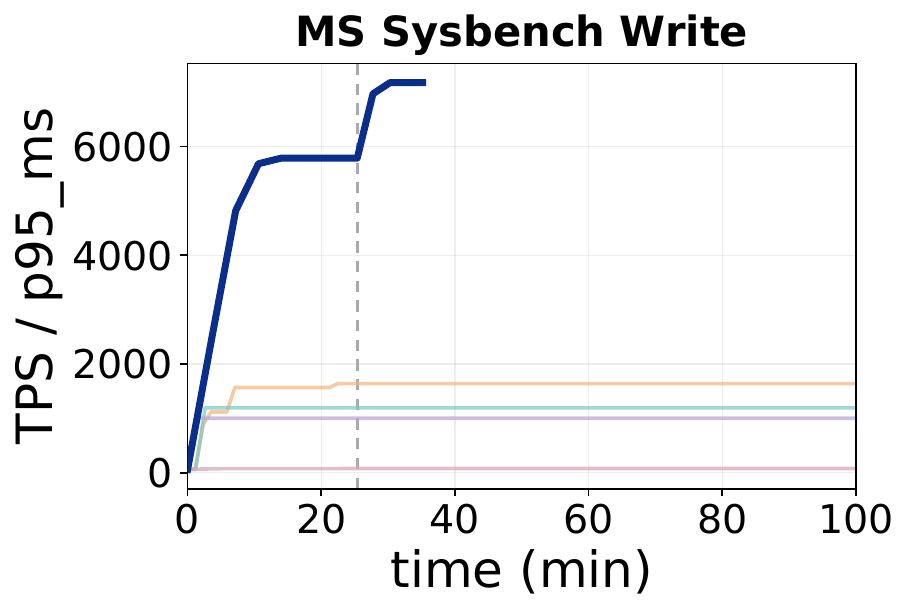}
        \includegraphics[width=0.32\linewidth]{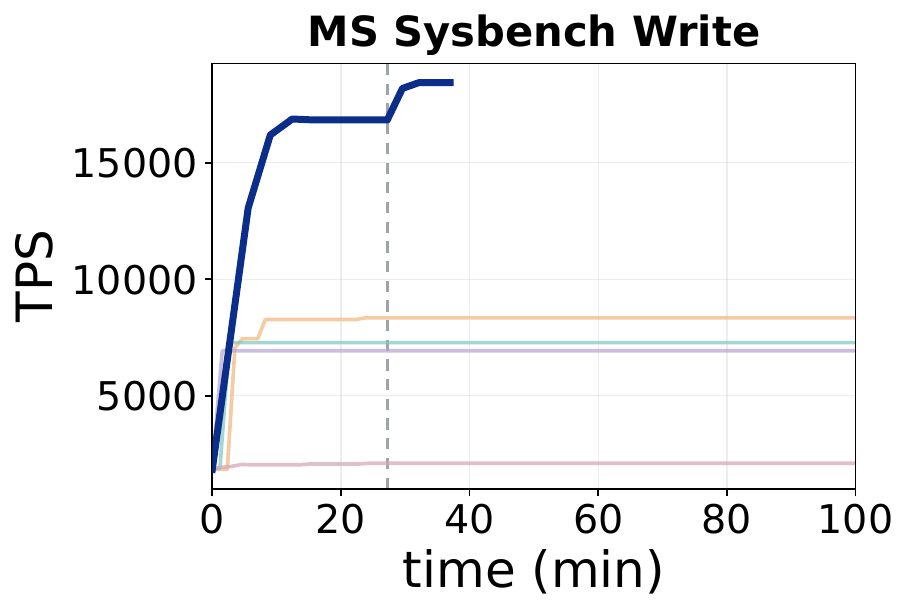}
        \includegraphics[width=0.32\linewidth]{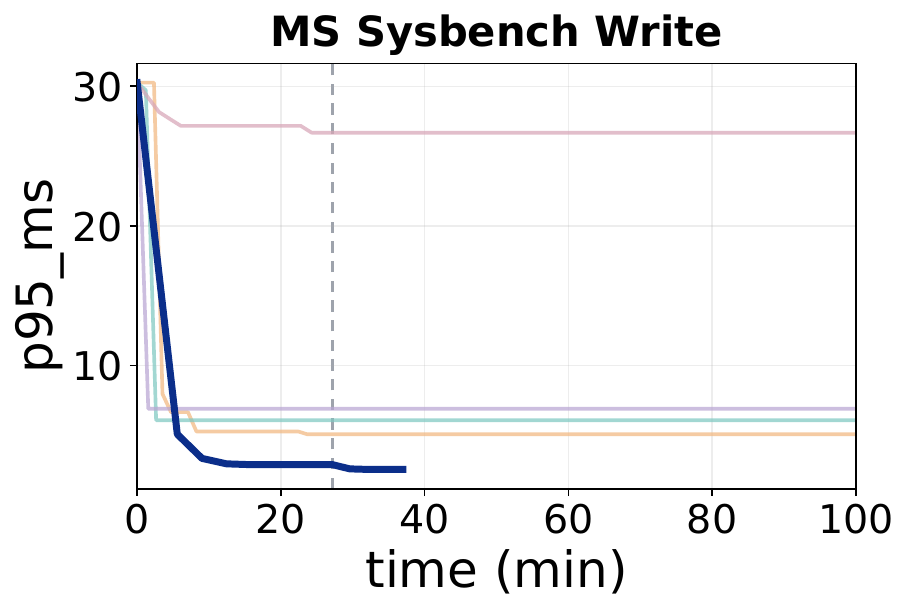}

        \vspace{1mm}

        % MS TPC-H
        \raggedright
        \includegraphics[width=0.32\linewidth]{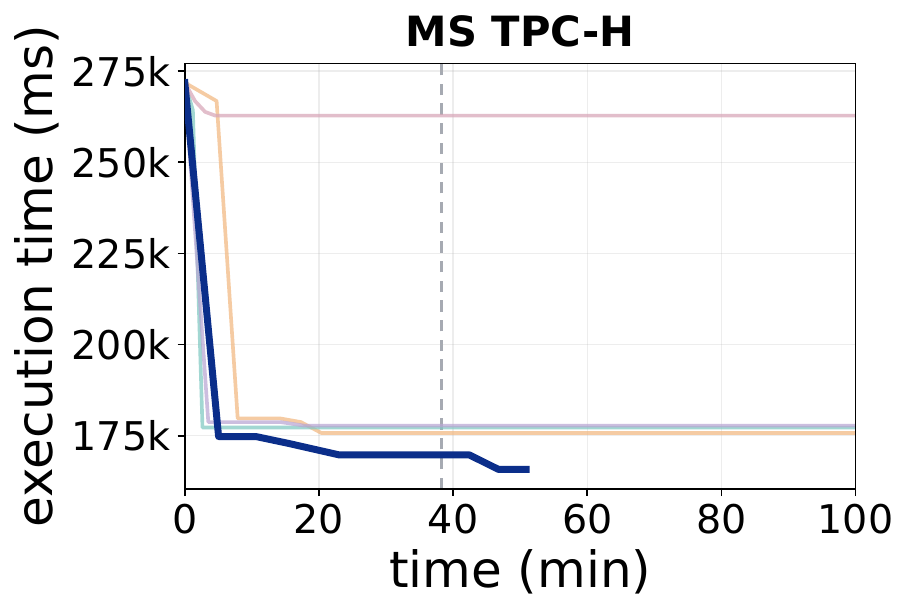}
        \hspace{0.32\linewidth}
        \hspace{0.32\linewidth}
    \end{minipage}
    \end{tabular}

    \captionsetup{skip=2pt}
    \caption{Overall reconfiguration progression of \textsc{AgenticDB} and
baseline methods across MySQL and PostgreSQL workloads. MS and PG
denote MySQL and PostgreSQL, respectively.}
    \label{fig:rq1-overall}
\vspace{-2mm}
\end{figure*}
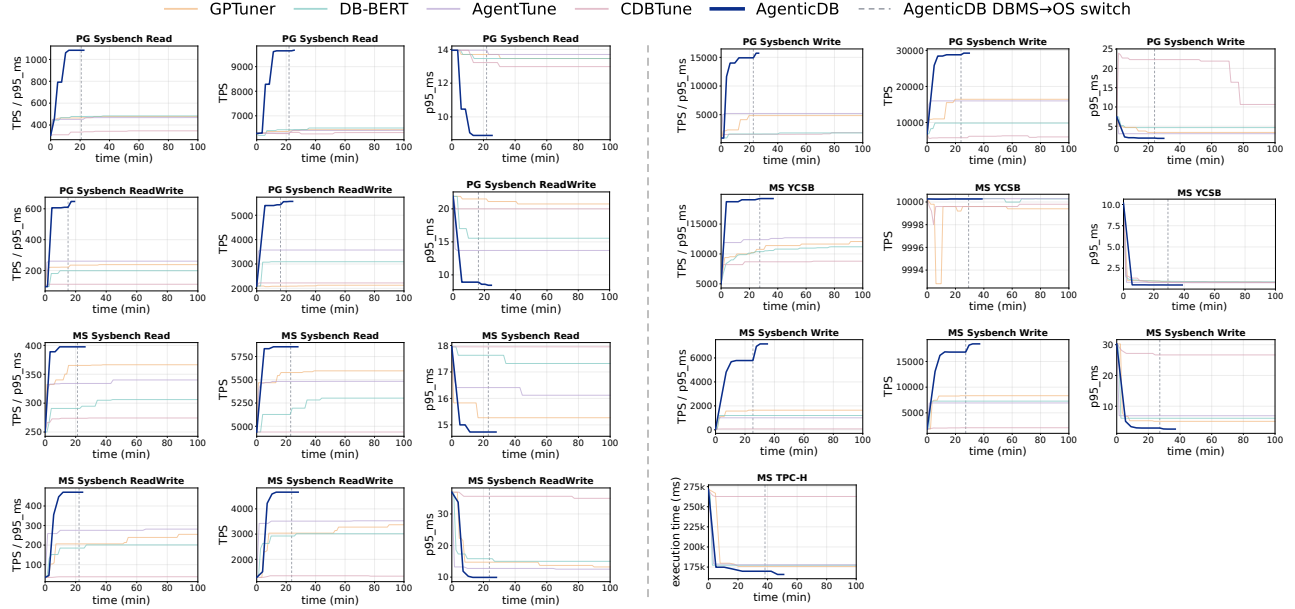

\begin{table*}[t]
\centering
\captionsetup{skip=2pt}
\caption{Best performance and time-to-best in RQ1. Entries report performance/time-to-best; time-to-best is in minutes; TPC-H improvement denotes execution-time reduction.}
\label{tab:rq1-all-methods}
\small
\begin{tabular}{lcccccc}
\toprule
\textbf{Workload}
& \textbf{GPTuner}
& \textbf{DB-BERT}
& \textbf{AgentTune}
& \textbf{CDBTune}
& \textbf{AgenticDB}
& \textbf{Improvement} \\
\midrule

MS YCSB
& 12076.6 / 91.5
& 11198.5 / 77.9
& \underline{12706.8 / 45.0}
& 8790.2 / 65.2
& \textbf{19262.2 / 27.3}
& \textbf{+51.6\%} (AgentTune) \\

MS Sysbench Read
& \underline{366.47 / 37.8}
& 306.16 / 44.2
& 340.16 / 44.5
& 274.30 / 28.8
& \textbf{397.48 / 9.5}
& \textbf{+8.5\%} (GPTuner) \\

MS Sysbench Write
& \underline{1638.4 / 22.4}
& 1195.6 / 2.6
& 1002.7 / 1.5
& 78.79 / 24.3
& \textbf{7171.9 / 30.3}
& \textbf{+337.7\%} (GPTuner) \\

MS Sysbench ReadWrite
& 254.97 / 87.5
& 200.96 / 27.0
& \underline{281.78 / 65.9}
& 38.14 / 77.3
& \textbf{470.45 / 11.7}
& \textbf{+67.0\%} (AgentTune) \\

MS TPC-H
& \underline{175734 / 20.4}
& 177239 / 18.4
& 177783 / 17.8
& 262758 / 4.5
& \textbf{165769 / 46.8}
& \textbf{+5.7\%} (GPTuner) \\

PG Sysbench Read
& 478.00 / 29.6
& \underline{483.20 / 35.6}
& 467.93 / 91.3
& 346.73 / 53.1
& \textbf{1083.6 / 12.4}
& \textbf{+124.3\%} (DB-BERT) \\

PG Sysbench Write
& 4838.6 / 19.7
& 1815.1 / 41.3
& \underline{5209.9 / 2.7}
& 1863.3 / 78.1
& \textbf{15724.7 / 24.7}
& \textbf{+201.8\%} (AgentTune) \\

PG Sysbench ReadWrite
& 238.42 / 35.5
& 198.92 / 9.8
& \underline{260.81 / 2.5}
& 111.14 / 1.5
& \textbf{647.61 / 17.0}
& \textbf{+148.3\%} (AgentTune) \\
\bottomrule
\end{tabular}
\vspace{-2mm}
\end{table*}

\subsection{RQ1: Overall Performance and Efficiency}
\label{sec:rq1-overall}

RQ1 evaluates the performance and tuning efficiency of
\textsc{AgenticDB} against representative database tuning baselines
across different workloads and DBMSs, including GPTuner, DB-BERT,
AgentTune, and CDBTune. Figure~\ref{fig:rq1-overall} presents the
performance progression over time, and
Table~\ref{tab:rq1-all-methods} summarizes the best performance,
time-to-best, and improvement over the best-performing baseline.

Figure~\ref{fig:rq1-overall} and Table~\ref{tab:rq1-all-methods} show that \textsc{AgenticDB} achieves the best performance across all evaluated workloads. The baselines exhibit different but limited
improvement patterns: AgentTune often obtains strong initial gains but
levels off after the first few rounds, GPTuner and DB-BERT improve more
gradually through iterative optimization, and CDBTune progresses more
slowly within the evaluation budget. In contrast, \textsc{AgenticDB}
combines rapid initial improvement with continued gains. Compared with
the best-performing baseline on each workload, it improves the objective
by 118.1\% on average, with gains ranging from 5.7\% to 337.7\%.
For OLTP workloads, the separate TPS and P95 latency curves show that
the improvement in $\mathrm{TPS}/\mathrm{P95}$ reflects both higher
throughput and lower tail latency. These gains are especially pronounced on write-intensive workloads,
reaching 337.7\% on MySQL Sysbench \texttt{write} and 201.8\% on
PostgreSQL Sysbench \texttt{write}. On MySQL TPC-H, \textsc{AgenticDB} reduces total query
execution time to 165.8 seconds, a 5.7\% reduction over the
best-performing baseline.

Figure~\ref{fig:rq1-overall} also shows that \textsc{AgenticDB} reaches
its best configuration 1.6$\times$--5.6$\times$ faster than the
best-performing baseline on several workloads. Across all workloads,
the best-performing baselines require 232.3 minutes in total to reach
their best configurations, whereas \textsc{AgenticDB} requires
179.7 minutes, reducing total time-to-best across workloads by 22.6\%.

\begin{table*}[t]
\centering
\captionsetup{skip=2pt}
\caption{Additional gains from OS-level reconfiguration after DBMS-level performance plateaus.}
\label{tab:rq2-os-gain}
\small
\begin{tabular}{lcccc}
\toprule
\textbf{Workload} & \textbf{Best before OS-level} & \textbf{Best after OS-level} & \textbf{OS Gain} & \textbf{Representative OS Parameter Groups} \\
\midrule
MS YCSB & 19060.2 & 19262.2 & +1.1\% 
& CPU scheduling, memory reclamation \\

MS Sysbench Write & 5784.2 & 7171.9 & +24.0\% 
& AIO limits, CPU scheduling, dirty-page writeback \\

MS TPC-H & 169756 ms & 165769 ms & +2.3\% 
& block I/O scheduler, read-ahead, THP control \\

PG Sysbench Write & 14918.8 & 15724.7 & +5.4\% 
& dirty-page writeback, CPU scheduling, memory reclamation \\

PG Sysbench ReadWrite & 610.03 & 647.61 & +6.2\% 
& dirty-page writeback, CPU scheduling, memory reclamation \\
\bottomrule
\end{tabular}
\vspace{-2mm}
\end{table*}

These results reflect the coupled effect of the two mechanisms in
\textsc{AgenticDB}. With a broad reconfiguration scope covering DBMS
knobs and OS parameters, context-grounded bottleneck diagnosis identifies
remaining bottlenecks from workload characteristics and runtime states
and translates the diagnosis into relevant reconfiguration actions.
Closed-loop context evolution further updates the planning context with
the outcomes of previous rounds, allowing subsequent diagnosis and
reconfiguration to focus on what remains unresolved rather than
repeatedly refining knob values within a limited DBMS knob space. As a
result, \textsc{AgenticDB} reaches a higher performance level after the
baselines have plateaued and reduces time-to-best across workloads.

\begin{answerbox}
\textbf{Answer to RQ1:}
\textsc{AgenticDB} achieves the best performance across all evaluated
workloads, outperforming the best-performing baseline by 118.1\% on
average. It also reduces total time-to-best from 232.3 to 179.7 minutes,
a 22.6\% reduction, achieving higher performance with less tuning time.
\end{answerbox}
% \vspace{-2mm}

\subsection{RQ2: Additional Improvement from OS-level Reconfiguration}
\label{sec:rq2-os}

RQ2 examines the additional improvement provided by OS-level
reconfiguration after DBMS-level performance plateaus. In
Figure~\ref{fig:rq1-overall}, the dashed vertical line marks the point
where \textsc{AgenticDB} switches from DBMS-level to OS-level
reconfiguration. We measure the OS-level gain by comparing the best
performance before this switch with the best performance obtained after
OS-level reconfiguration starts.

Table~\ref{tab:rq2-os-gain} reports the workloads for which OS-level
reconfiguration yields a positive additional gain. The largest
improvement occurs on MySQL Sysbench \texttt{write}, where the objective
increases from 5784.2 to 7171.9, corresponding to a 24.0\% gain.
PostgreSQL Sysbench \texttt{readwrite} and \texttt{write} improve by
6.2\% and 5.4\%, respectively, while MySQL YCSB and TPC-H obtain
smaller gains of 1.1\% and 2.3\%. The remaining workloads show no
further improvement after the switch to OS-level reconfiguration.

These results show that OS-level reconfiguration can provide additional
improvement when DBMS-level reconfiguration leaves residual bottlenecks
outside DBMS knobs. For example, write-intensive workloads may still be
affected by OS-level CPU scheduling, asynchronous I/O capacity, memory
reclamation, or writeback behavior, making the corresponding OS
parameters useful in later rounds. When the remaining bottlenecks are less sensitive to OS parameters,
OS-level changes provide limited additional benefit. \textsc{AgenticDB} therefore enters OS-level reconfiguration after
DBMS-level performance plateaus, explores whether OS parameters can
further improve performance, and terminates when no further improvement is observed.

\begin{answerbox}
\textbf{Answer to RQ2:}
OS-level reconfiguration provides up to 24.0\% additional improvement
after DBMS-level performance plateaus, demonstrating the benefit of
extending reconfiguration beyond DBMS parameters.
\end{answerbox}

\subsection{RQ3: Reliability of the Reconfiguration Process}
\label{sec:rq3-reliability}

RQ3 evaluates how the Validator and Recovery improve the reliability of
the reconfiguration process. The Validator checks proposed
configurations before they are applied, whereas Recovery handles
failures reported by the Executor during configuration application or workload execution.
Table~\ref{tab:rq3-robustness-cases} summarizes three representative
cases from the reconfiguration traces.

\begin{table*}[t]
\centering
\captionsetup{skip=2pt}
\caption{Representative validation and recovery cases in \textsc{AgenticDB}.}
\label{tab:rq3-robustness-cases}
\footnotesize
\setlength{\tabcolsep}{3pt}
\begin{tabularx}{\textwidth}{
@{}
>{\raggedright\arraybackslash}p{0.085\textwidth}
>{\raggedright\arraybackslash}p{0.14\textwidth}
>{\raggedright\arraybackslash}X
>{\raggedright\arraybackslash}X
@{}
}
\toprule
\textbf{Component} & \textbf{Case} & \textbf{Detected problem} & \textbf{Outcome} \\
\midrule
Validator
& MySQL Sysbench \texttt{write}
& Supported but semantics-changing setting:
  \path|transaction_isolation=READ-COMMITTED|
& Rejected before execution; avoids an invalid benchmark trial \\

Validator
& OS-level reconfiguration
& Unsupported sysctl parameter:
  \path|kernel.sched_migration_cost_ns|
& Rejected before execution; revised using validation feedback \\

Recovery
& PostgreSQL restart
& Inconsistent \path|wal_level=minimal| with enabled WAL sender settings
& Restores a compatible WAL setting and DBMS availability \\
\bottomrule
\end{tabularx}
\vspace{-2mm}
\end{table*}

\paragraph{Validator.}
The first two rows of Table~\ref{tab:rq3-robustness-cases} show how the
Validator prevents invalid configurations before execution. In
the MySQL Sysbench \texttt{write} case, the LLM DBA Planner proposes
\path|transaction_isolation=READ-COMMITTED|. Although the parameter is
supported by MySQL, applying it would change the transaction semantics
of the benchmark rather than only improve system performance. The
Validator rejects this configuration and returns structured feedback to
the LLM DBA Planner for revision.

In the OS-level case, the LLM DBA Planner proposes the unsupported sysctl
parameter \path|kernel.sched_migration_cost_ns|. The Validator detects
the unknown entry before execution, preventing an unsupported OS
parameter from reaching the Executor.

\begin{table}[t]
\centering
\captionsetup{skip=2pt}
\caption{Timing of a semantics-changing DBMS candidate with and without the Validator.}
\label{tab:rq3-validator-timing}
\small
\begin{tabular}{lcc}
\toprule
\textbf{Step} & \textbf{With Validator} & \textbf{Without Validator} \\
\midrule
LLM proposal & 42.69 s & 42.69 s \\
Validation check & $<0.01$ s & 0 s \\
Apply configuration & 0 s & 1.87 s \\
Warmup + benchmark & 0 s & 84.31 s \\
Rollback & 0 s & 3.68 s \\
\midrule
Total time & $<42.70$ s & 132.55 s \\
Avoided downstream time & -- & 89.86 s \\
\bottomrule
\end{tabular}
\vspace{-2mm}
\end{table}

Table~\ref{tab:rq3-validator-timing} quantifies the cost avoided in the
semantics-changing case. The deterministic local validation check takes
less than 0.01 seconds. Without validation, the proposed configuration
would proceed through configuration application, benchmark execution,
and rollback, introducing 89.86 seconds of avoidable time.

\paragraph{Recovery.}
Recovery handles failures that cannot be fully identified before
execution. In the PostgreSQL case, the LLM DBA Planner proposes a
configuration that sets \path|wal_level=minimal| while WAL sender
settings such as \path|max_wal_senders| remain enabled. The resulting
dependency conflict is exposed during restart and prevents PostgreSQL
from becoming available. Recovery inspects the service status and error
logs, identifies the conflicting WAL setting, restores
\path|wal_level=replica|, restarts PostgreSQL, and verifies service
availability. The failure and recovery outcome are recorded in session
memory so that later rounds can avoid the same configuration pattern.
Without this recovery path, PostgreSQL would fail to restart, and the
reconfiguration process would require manual repair before it could
continue.

\begin{answerbox}
\textbf{Answer to RQ3:}
The Validator and Recovery improve the reliability of the
reconfiguration process. The Validator rejects invalid configurations
before execution, while Recovery handles execution failures through
repair or rollback, avoiding manual repair.
\end{answerbox}

\subsection{RQ4: Effects of LLM Backends and Experience Memory}
\label{sec:rq4-llm-memory}

RQ4 examines two design choices in \textsc{AgenticDB}: the LLM backend
and experience memory. We first compare multiple LLM backends under the same
\textsc{AgenticDB} framework to assess their effects on
reconfiguration performance, total tuning time, and API cost. We then evaluate whether experience memory provides additional
benefit beyond session memory.

\paragraph{Effect of LLM backends.}
We compare five LLM backends: GPT-5.5, DeepSeek V4 Pro, Claude Opus 4.7, GLM 5.1, and Kimi K2.6~\cite{openai-gpt55,deepseek-v4-report,
claude-opus47,glm51,kimi-k26}.

\begin{figure}[t]
\vspace{-2mm}
    \centering

    \includegraphics[width=0.85\columnwidth]{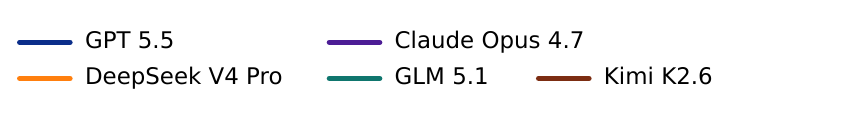}

    \vspace{-1mm}

    \includegraphics[width=0.48\columnwidth]{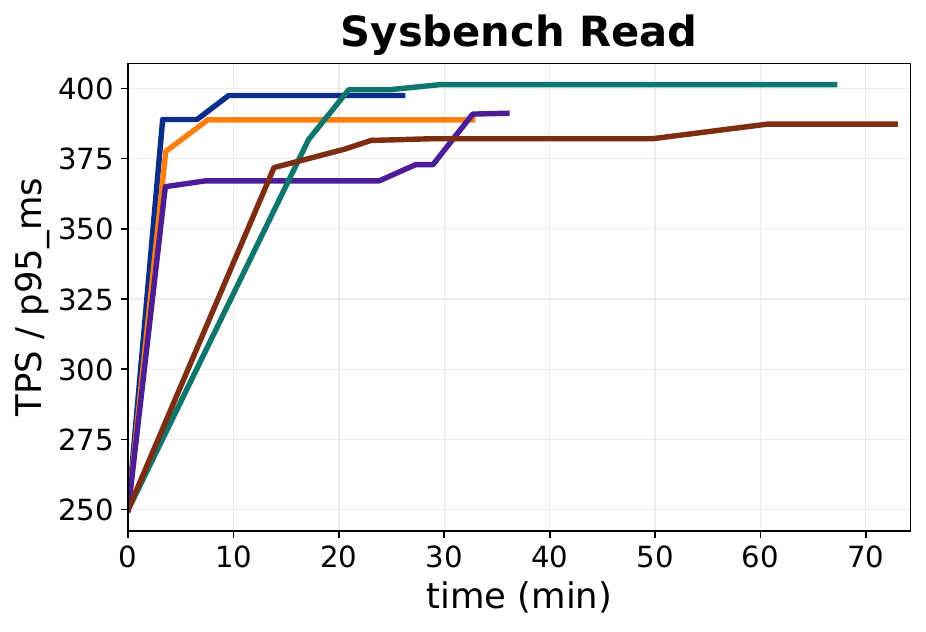}
    \includegraphics[width=0.48\columnwidth]{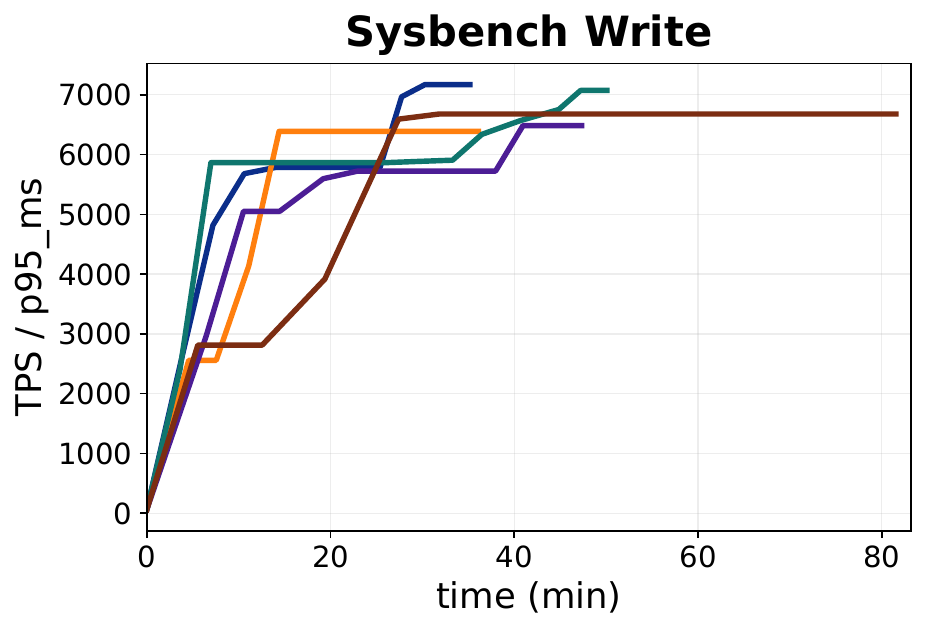}

    \vspace{-1mm}

    \includegraphics[width=0.48\columnwidth]{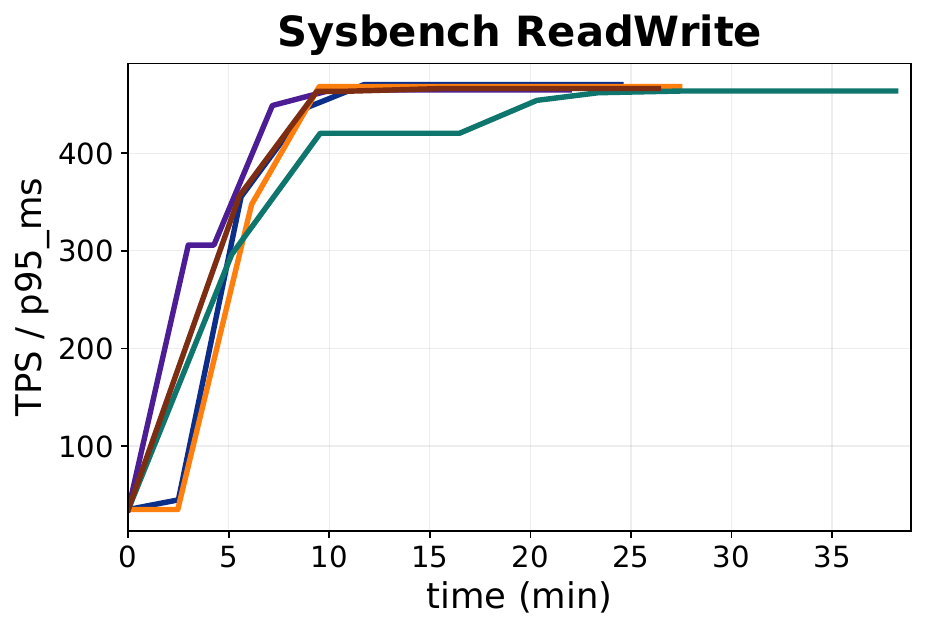}
    \includegraphics[width=0.48\columnwidth]{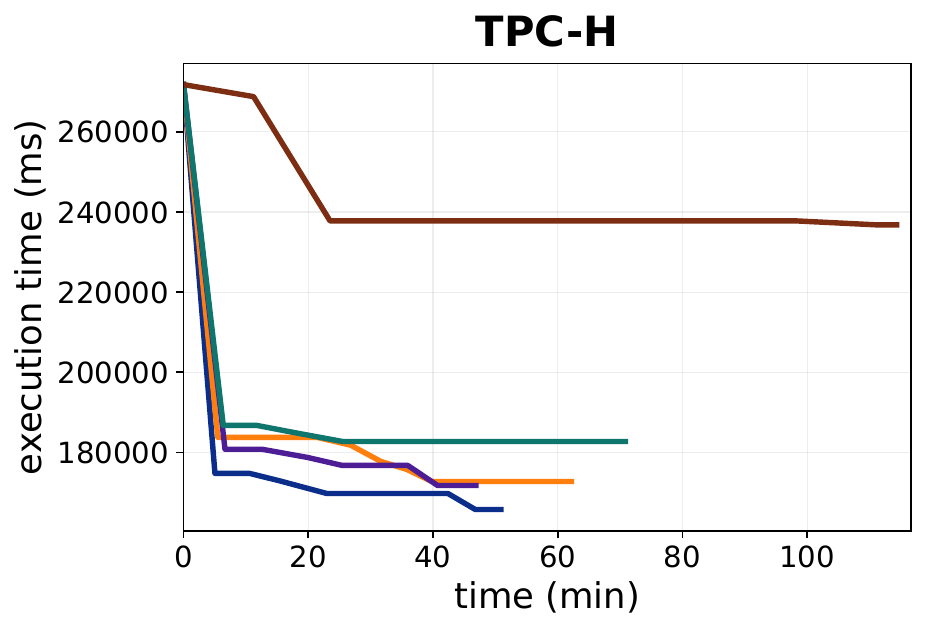}

    \captionsetup{skip=2pt}
    \caption{
    Reconfiguration progression of \textsc{AgenticDB} with different LLM backends on representative MySQL workloads.
    }
    \label{fig:rq4-llm-backends}
\vspace{-2mm}
\end{figure}

\begin{table}[t]
\centering
\captionsetup{skip=2pt}
\caption{Average LLM calls, token usage, and API cost per workload. Avg. LLM Calls denotes the average number of Planner invocations.}
\label{tab:rq4-token-cost}

\begingroup
\small
\renewcommand{\arraystretch}{0.88}
\setlength{\tabcolsep}{4pt}
\setlength{\aboverulesep}{0.20ex}
\setlength{\belowrulesep}{0.20ex}

\begin{tabular}{@{}lrrr@{}}
\toprule
\textbf{Model}
& \textbf{Avg. LLM Calls}
& \textbf{Avg. Tokens}
& \textbf{Avg. API Cost} \\
\midrule
GPT-5.5          & 11.00 & 584,458 & \$3.97 \\
DeepSeek V4 Pro  & 11.00 & 548,039 & \$0.26 \\
Claude Opus 4.7  & 10.75 & 795,192 & \$4.80 \\
GLM 5.1          & 12.50 & 696,826 & \$0.84 \\
Kimi K2.6        & 10.25 & 595,413 & \$0.68 \\
\bottomrule
\end{tabular}
\endgroup

% \vspace{-4mm}
\end{table}

Figure~\ref{fig:rq4-llm-backends} shows that \textsc{AgenticDB} can
operate with all evaluated LLM backends, while different backends
exhibit distinct trade-offs in reconfiguration performance, total tuning time, and API cost. GPT-5.5, Claude Opus 4.7, and DeepSeek V4 Pro complete reconfiguration
within relatively short tuning time, and GPT-5.5 usually has the
shortest tuning time. GPT-5.5 achieves the best overall performance among the evaluated
backends. It reaches configurations with high
objective values quickly on the Sysbench workloads and obtains the
lowest execution time on TPC-H. As shown in
Table~\ref{tab:rq4-token-cost}, GPT-5.5 costs \$3.97 per workload on
average. Claude Opus 4.7 also reaches competitive configurations on
several workloads, but does not outperform GPT-5.5 and incurs the
highest average API cost of \$4.80.

DeepSeek V4 Pro provides the most cost-effective alternative among the
evaluated backends. It achieves consistently strong reconfiguration
performance at an average cost of only \$0.26 per workload, although
its final performance remains below GPT-5.5. It therefore provides a
practical balance between reconfiguration performance, tuning time, and API cost.

GLM 5.1 and Kimi K2.6 also have relatively low API costs, but require
longer tuning time because of response latency and
API instability. GLM 5.1 occasionally reaches competitive
configurations but requires more LLM calls to complete the
reconfiguration process. Kimi K2.6 is particularly affected on TPC-H,
requiring substantially longer tuning time while still
producing a configuration with higher query execution time. This
observation is consistent with the DeepSeek-V4 technical report, which
reports unavailable evaluation entries for K2.6 and GLM-5.1 because
their APIs did not return responses reliably~\cite{deepseek-v4-report}.
Overall, GPT-5.5 provides the best overall reconfiguration performance,
while DeepSeek V4 Pro offers the most cost-effective alternative.

\begin{figure}[t]
    \centering

    \includegraphics[width=0.8\columnwidth]{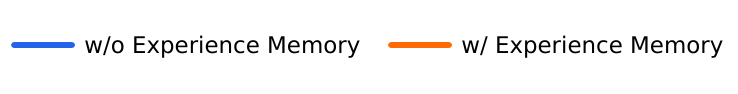}

    \vspace{-1mm}

    \includegraphics[width=0.45\columnwidth]{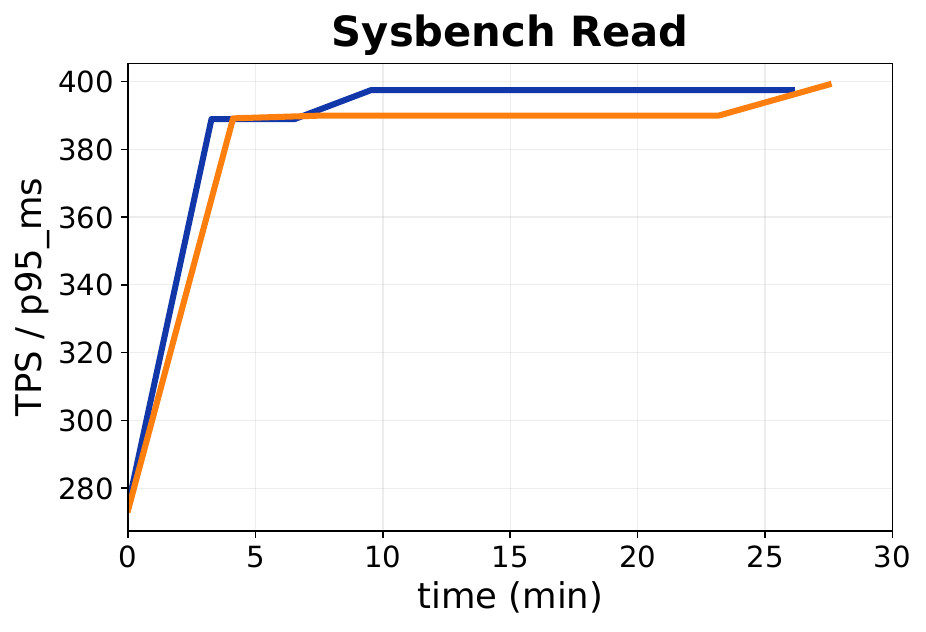}
    \includegraphics[width=0.45\columnwidth]{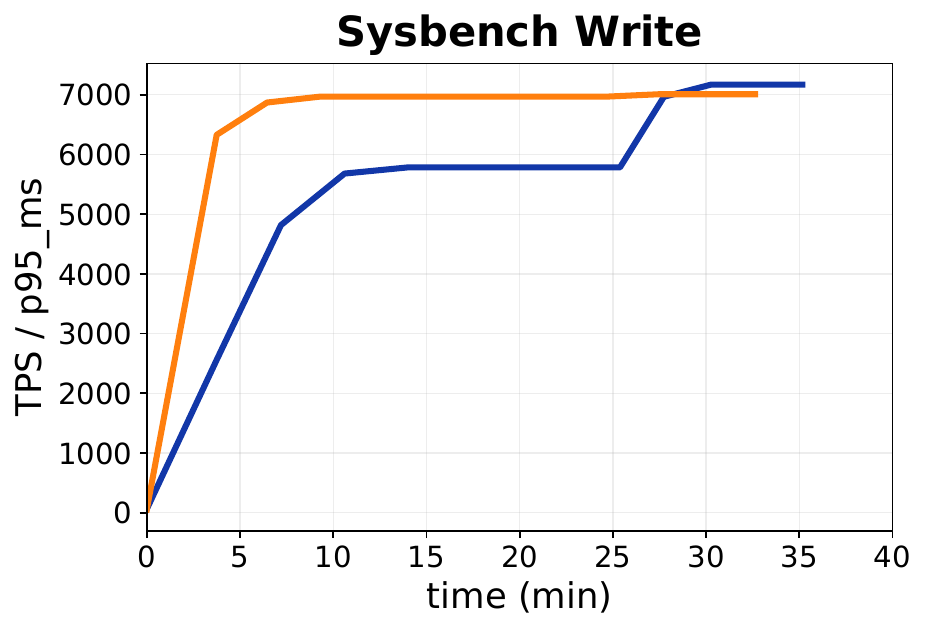}

    \vspace{0mm}

    \includegraphics[width=0.45\columnwidth]{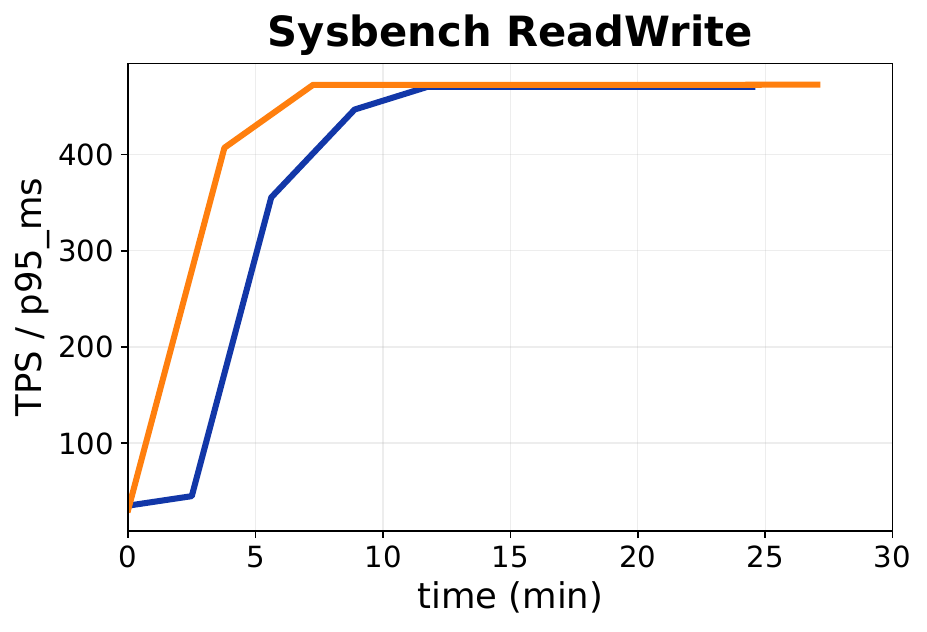}
    \includegraphics[width=0.45\columnwidth]{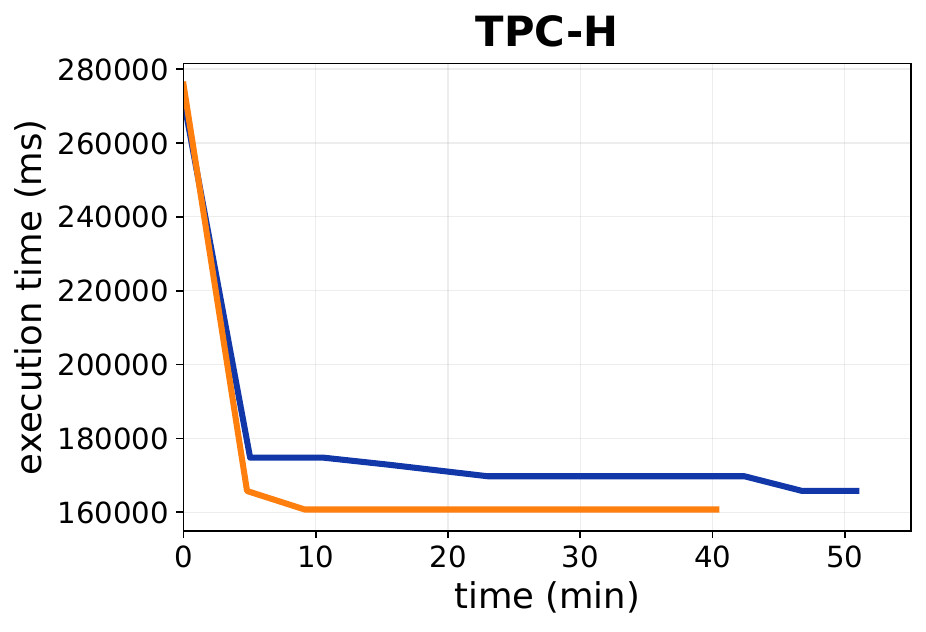}

    \captionsetup{skip=2pt}
    \caption{Performance of \textsc{AgenticDB} without and with experience memory on representative MySQL workloads.}
    \label{fig:rq4-memory}
    \vspace{-2mm}
\end{figure}

\paragraph{Effect of experience memory.}
Under the same GPT-5.5 backend, we compare \textsc{AgenticDB} with
session memory only and \textsc{AgenticDB} with both session memory and
experience memory. Figure~\ref{fig:rq4-memory} presents the results.

Experience memory improves reconfiguration mainly by providing prior
guidance at the beginning of a new reconfiguration process. The benefit
is most visible on Sysbench \texttt{write}, Sysbench
\texttt{readwrite}, and TPC-H. On Sysbench \texttt{read}, experience
memory does not bring a clear additional gain, because both variants
quickly reach a similar high-performing configuration for this
read-intensive workload. On the write-intensive and mixed
workloads, retrieved experience helps the LLM DBA Planner identify
promising reconfiguration directions earlier, reducing low-value trials
in the initial rounds. On TPC-H, it guides the process toward a
configuration region with lower execution time early in reconfiguration,
leaving later rounds to perform more focused refinement.

Overall, experience memory is beneficial when prior evidence provides a
useful initial direction, helping \textsc{AgenticDB} reach
high-performing configurations earlier in the reconfiguration process.

\begin{answerbox}
\textbf{Answer to RQ4:}
LLM backends differ in reconfiguration performance, total tuning time,
and API cost. GPT-5.5 achieves the best overall performance, while
DeepSeek V4 Pro offers the best cost-effectiveness. Experience memory
provides useful initial guidance, helping \textsc{AgenticDB} reach
high-performing configurations earlier.
\end{answerbox}

\section{Related Work}
\label{sec:related}

Database configuration tuning has been widely studied because DBMS
performance is highly sensitive to configuration choices
~\cite{tuning_inquiry,self_driving_dbms}. Prior methods can be broadly
categorized as search- and learning-based tuning,
historical-data-driven tuning, and knowledge- or
language-model-assisted tuning~\cite{cloud_tuning_survey,tuning_inquiry}.

\paragraph{Search- and learning-based tuning.}
Search-based methods explore predefined DBMS knob spaces using
heuristic search or surrogate models. BestConfig adopts heuristic
search~\cite{bestconfig}, while iTuned and SMAC formulate tuning as
black-box Bayesian optimization~\cite{ituned,smac}. Reinforcement-
learning systems such as CDBTune and QTune optimize configurations
through sequential interaction with DBMS metrics and workload
feedback~\cite{cdbtune,qtune}. HUNTER further combines warm-start
exploration, search-space reduction, and reinforcement
learning~\cite{hunter}. Despite reducing manual effort, these methods still require repeated
workload executions within predefined DBMS knob spaces, and their
feedback is mainly used to guide configuration search rather than to
explicitly reason about remaining bottlenecks.

\paragraph{Historical-data-driven tuning.}
Historical-data-driven methods transfer prior tuning experience to new
workloads. OtterTune identifies important knobs and recommends
configurations by matching new workloads with historical ones
~\cite{ottertune}. ResTune applies meta-learning across tasks under SLA
constraints~\cite{restune}, while E2ETune learns from historical
workload--configuration pairs~\cite{e2etune}. These methods can improve
tuning efficiency when relevant historical data are available, but their
effectiveness depends on the relevance and coverage of the historical
workload and configuration records. Their recommendations may also
become less reliable when the target hardware or deployment environment
differs from the historical data.

\paragraph{Knowledge- and language-model-assisted tuning.}
Recent methods incorporate domain knowledge and language models into
database tuning. DB-BERT extracts tuning knowledge from manuals and
combines it with reinforcement learning~\cite{dbbert}. GPTuner uses
LLM-derived knowledge for knob selection, search-space construction, and
coarse-to-fine Bayesian optimization~\cite{gptuner}. LlamaTune improves
sample efficiency through domain-aware dimensionality reduction and
value bucketization~\cite{llamatune}. $\lambda$-Tune generates complete
configuration scripts with LLMs and evaluates selected candidates
~\cite{lambdatune}. AgentTune decomposes DBMS knob tuning into workload
analysis, knob selection, range pruning, configuration recommendation,
and feedback-driven refinement~\cite{agenttune}. These methods reduce the tuning cost of LLM-assisted optimization, but
they often do so by narrowing the configuration space through knob
selection, range pruning, or value bucketization. As a result, many
potentially useful configurations may remain unexplored.

\paragraph{Summary.}
Prior methods have advanced automatic tuning through search, learning,
historical experience reuse, and language-model guidance. However, they
can still leave substantial performance potential unexplored, suffer
from low tuning efficiency, and provide limited support for validating
proposed changes, handling reload or restart requirements, and
recovering from execution failures. \textsc{AgenticDB} instead
organizes tuning as a closed-loop DBMS/OS reconfiguration process with
context-grounded bottleneck diagnosis, closed-loop context evolution,
configuration validation, and failure recovery.

\section{Conclusion}
\label{sec:conclusion}

This paper presents \textsc{AgenticDB}, a self-evolving agentic
framework for database workload reconfiguration. \textsc{AgenticDB} organizes tuning as a closed-loop DBMS/OS reconfiguration
process that uses runtime states and execution feedback to diagnose
bottlenecks, update the planning context, recommend configuration
actions, validate them before application, and recover from failures. Experiments on MySQL and PostgreSQL with
OLTP and OLAP workloads show that \textsc{AgenticDB} achieves the best
performance across all evaluated workloads, improving over the
best-performing baseline by 118.1\% on average and reducing total
time-to-best across workloads by 22.6\%.

% \begin{acks}
% The work was supported by the Strategic Priority Research Program of the Chinese Academy of Sciences under Grant No.~XDA0360202.
% \end{acks}

%\clearpage

\bibliographystyle{ACM-Reference-Format}
\bibliography{sample}

\end{document}